\documentclass[twocolumn,citeautoscript,pra,superscriptaddress,amsmath,amssymb]{revtex4-1}
 
\usepackage{graphicx}
\usepackage{xcolor}
\usepackage{upgreek}
\usepackage{float}
\usepackage{lipsum}
\usepackage{mathrsfs}
\usepackage{booktabs}
\usepackage{url}
\usepackage[colorlinks,citecolor=blue,linkcolor=blue]{hyperref}
\usepackage{amsmath}
\usepackage{amsthm}
\usepackage{mathtools}
\usepackage{multirow}
\usepackage{algorithm} 
\usepackage{algpseudocode}
\usepackage{bbold}
\usepackage[export]{adjustbox}

\theoremstyle{definition}
\newtheorem{result}{Result}

\newtheorem{definition}{Definition}

\newcommand{\ket}[1]{\ensuremath{\left| #1 \right\rangle}}
\newcommand{\bra}[1]{\ensuremath{\left\langle #1 \right|}}

\begin{document}

\title{On the sampling complexity of open quantum systems}
\author{I. A. Aloisio}
\email{isobel.aloisio@monash.edu}
\affiliation{School of Physics and Astronomy, Monash University, Clayton, VIC 3800, Australia}

\author{G. A. L. White}
\email{white.g@unimelb.edu.au}
\affiliation{School of Physics, University of Melbourne, Parkville, VIC 3010, Australia}

\author{C. D. Hill}
\email{cdhill@unimelb.edu.au}
\affiliation{School of Physics, University of Melbourne, Parkville, VIC 3010, Australia}
\affiliation{School of Mathematics and Statistics, University of Melbourne, Parkville, VIC, 3010, Australia}

\author{K. Modi}
\email{kavan.modi@monash.edu}
\affiliation{School of Physics and Astronomy, Monash University, Clayton, VIC 3800, Australia}
\affiliation{Centre for Quantum Technology, Transport for New South Wales, Sydney, NSW 2000, Australia}

\begin{abstract}
Open quantum systems are ubiquitous in the physical sciences, with widespread applications in the areas of chemistry, condensed matter physics, material science, optics, and many more. Not surprisingly, there is significant interest in their efficient simulation. However, direct classical simulation quickly becomes intractable with coupling to an environment whose effective dimension grows exponentially. This raises the question: can quantum computers help model these complex dynamics? A first step in answering this question requires understanding the computational complexity of this task. Here, we map the temporal complexity of a process to the spatial complexity of a many-body state using a computational model known as the process tensor framework. With this, we are able to explore the simulation complexity of an open quantum system as a dynamic sampling problem: a system coupled to an environment can be probed at successive points in time -- accessing multi-time correlations. The complexity of multi-time sampling, which is an important and interesting problem in its own right, contains the complexity of master equations and stochastic maps as a special case. Our results show how the complexity of the underlying quantum stochastic process corresponds to the complexity of the associated family of master equations for the dynamics. We present both analytical and numerical examples whose multi-time sampling is as complex as sampling from a many-body state that is classically hard. This also implies that the corresponding family of master equations are classically hard. Our results pave the way for studying open quantum systems from a complexity-theoretic perspective, highlighting the role quantum computers will play in our understanding of quantum dynamics.
\end{abstract}
\maketitle




\section{Introduction}
With the rapid improvement in quantum computing technologies, there is increasing interest in finding practical problems that will demonstrate a quantum advantage over classical methods. A leading candidate is the simulation of strongly correlated quantum systems~\cite{mcclean2021foundations}. 
A compelling subclass of these dynamics are open quantum systems whose correlations lie between an accessible system and an inaccessible environment. These constitute some of the most challenging yet fascinating physical phenomena, and underlie the fields of quantum chemistry, materials science, and condensed matter physics~\cite{smith2019simulating,cao2019quantum, mcardle2020quantum, cohen2011memory, kandala2017hardware}. Contexts range from the study of strongly correlated materials such as glass-forming liquids~\cite{kim2010multi}, to conformational dynamics of proteins~\cite{ono2015couplings}, to understanding the efficiency of photosynthesis~\cite{mazziotti2012effect}. \par
Understanding the dynamics of open quantum systems -- in terms of master equations, stochastic maps, or their multi-time correlations -- requires additional considerations over closed ones. Naively, classical simulation of a small system may seem tractable. But accounting for the relevant effects of environmental coupling on system evolution increases the problem manifold. 
From the opposite angle, one might ask whether the simulation of an open quantum system is as difficult as solving the entire dilated dynamics. But this approach is asking too much: the most compressed description of a system can often be ignorant to the specifics of the environment. 
The true difficulty of the problem lies somewhere in between. Coherent system-environment interactions can lead to an exchange of information between the two, and generate highly complex temporal correlations across the system dynamics. In other words, it is the complexity of the memory -- or the non-Markovianity -- which governs the difficulty of the problem. 
At one end of the spectrum, memoryless dynamics are almost as easy to solve as the closed dynamics of the system. At the other extreme, the evolution of the system can be every bit as complex as the system and its environment.
Precisely identifying, defining, and exploring this transition is the purpose of this work.

So then, exactly how complex are non-Markovian processes from a complexity-theoretic perspective? How hard is it to classically simulate a master equation, a stochastic map, or multi-time correlations? Answering these questions will inform which physical problems may be good candidates for simulating on quantum computers and which are solvable with classical methods. 
Moreover, this will inform as to how quantum simulations should be designed on real quantum devices, to extract meaningful information from complex quantum processes~\cite{li2021succinct, cleve2017}.
And more fundamentally, it is the key to determining if multi-time correlations of a single qubit can be complex, if a noisy process can be complex, and consequently if we will see any computational advantage using noisy intermediate-scale quantum (NISQ) devices.


The efforts to model the non-Markovian quantum dynamics on classical computers are vast and the \textit{simulation complexity} of many models is well-studied. However, typically the complexity of classically simulating open quantum systems is attributed to the difficulty of solving the closed dynamics
of the system and its environment, which we refer to as \textit{dilated complexity}. 
In this setting, the computational cost of direct simulation scales exponentially in the size of the environment. Numerous classical techniques exist to tackle the simulation of these complex environments~\cite{de2017dynamics,rosenbach2016efficient}, including automated environment compression~\cite{cygorek2022simulation}, time evolving matrix product operators~\cite{strathearn2018efficient}, path-integral based approximations~\cite{breuer2002theory,strathearn2017efficient,jorgensen2019exploiting}, the ensemble of Lindblad's trajectories~\cite{head2021capturing}, and quantum Monte carlo~\cite{nagy2019variational, chen2017inchworm, cohen2013numerically}.

Specific dimension-based metrics have been proposed to quantify the simulation complexity. This includes the transfer tensor method~\cite{cerrillo2014non, gherardini2022transfer, buser2017initial, kananenka2016accurate}, the matrix bond dimension of the minimum effective environment~\cite{luchnikov2019simulation}, and the rank of the time delay matrix~\cite{khan2021model}. 
Yet, their relation to specific physics models remains tenuous. 
It is important to note that the dilated complexity will be always greater than or equal to that of the actual simulation complexity. 
In other words, the latter should not be burdened by the cumulative task of incorporating all environmental degrees of freedom and then subsequently reducing the problem back to the open dynamics by means of a partial trace. 
On the other hand, the classical algorithms listed above are not always accurate with long-time dynamics or strong system-environment coupling~\cite{jung2020ring}. 
This suggests that the simulation complexity may indeed be on par with the dilated complexity in these settings.


Our approach will be to think of simulating open quantum dynamics, which includes master equations, stochastic maps, and multi-time correlations, as a quantum sampling problem. 
Indeed, from a complexity perspective, sampling complexity is used as a definition for simulation complexity, since it replicates what is obtainable in an experiment where a quantum state is prepared, evolved for some fixed time, and then measured~\cite{lund2017quantum}. 
This perspective has led to several key proposals for quantum supremacy, including BosonSampling~\cite{aaronson2011computational}, $\mathsf{IQP}$~\cite{bremner2011classical}, and Random Circuit Sampling (RCS)~\cite{boixo2018characterizing}, as well as more physical models such as driven many-body systems~\cite{tangpanitanon2020quantum}, Ising spin models~\cite{fefferman2017exact} and bosonic lattice models~\cite{deshpande2018dynamical}.
However, attempts to incorporate the effects of interaction with an uncontrollable environment on the sampling complexity of a quantum system have been restricted to Markovian, or memoryless, processes~\cite{kapit2020entanglement, shtanko2021complexity}. 
Once again, the conventional sampling problem considers the dilated simulation complexity of a closed system. Thus, our key proposition for open quantum dynamics is to sample a single system at many consecutive times. The task therefore sheds the environmental baggage and distills the question of open quantum system complexity solely to a single controllable system. The presence of complex temporal correlations directly translates to the complexity of this task.
We formalise this approach using a computational model known as the process tensor framework, a complete mathematical description for quantum stochastic processes~\cite{pollock2018non}.

Our work is structured as follows. We begin in Section~\ref{section:Computational model} by describing the process tensor framework in detail and argue why it is an ideal computational model for defining the task of sampling an open quantum system over time. Using our model we show that dynamic multi-time sampling of an open quantum system has an equivalent representation as a sampling problem on a many-body state, formalising a relationship between the complexity of states and the complexity of processes. We then argue that the complexity of the underlying quantum stochastic process is comparable to the complexity of the associated family of master equations. In Section~\ref{section:OpenDQP} we define the complexity class \textit{Open Dynamic Quantum Polynomial-time} ($\mathsf{OpenDQP}$), which is the class of problems solvable within our model, and use its relationship to other complexity classes to classify how complex an open quantum system can be. This complexity class is then examined under several sampling models, including the Heisenberg interaction in a spin chain, $\mathsf{IQP}$, and random processes, where we identify instances of $\mathsf{OpenDQP}$ which exist in $\mathsf{BQP} \setminus \mathsf{BPP}$ and are thus hard to classically sample from. The backslash denotes the relative complement of $\mathsf{BPP}$ with respect to $\mathsf{OpenDQP}$.
Ultimately, our results clarify possible avenues to quantum advantage in simulating open quantum systems.

\section{Computational model for open quantum dynamics}
\label{section:Computational model}
We begin with outlining the type of problems one usually faces in the area of open quantum dynamics. Most open dynamics problems either involve quantum master equations or stochastic maps (also known as quantum channels or completely positive maps). Master equations are ubiquitous in chemistry, condensed matter physics, material science, quantum optics, and statistical physics~\cite{tanimura2006stochastic, zhang2019exact, brian2021generalized}. It turns out that, from a complexity theory perspective, the complexity of master equations is the same as the complexity of stochastic maps (we will elaborate on this below). However, in general, it is difficult to grapple with the computational complexity of these objects. Thus, our approach will be to focus on the computational complexity of the underlying process as described by high-order maps known as process tensors~\cite{pollock2018operational, pollock2018non, costa2016quantum}. The latter constitute the most general object in the theory of quantum stochastic processes, and thus contains both the master equations and stochastic maps as limiting cases. The process tensor, in its own right, is also of significant interest for understanding and controlling noise in modern quantum devices~\cite{milz2021quantum, white2020demonstration,white2021many,white2022non}.

We will start here with stating the most general master equation first, the \textit{Nakajima-Zwanzig master equation}.
\begin{gather}\label{eq:me}
    \partial_t \rho_t = \stackrel{\text{local driving}}{\overbrace{-i[H_t,\rho_t]}}
     + 
    \stackrel{\text{memory kernel}}{\overbrace{\int_0^t ds \, \mathcal{K}_s [\rho_s]}}
     + 
    \stackrel{\text{i.c.}}{\overbrace{\mathcal{J}_0}}.
\end{gather}
Above, the complexity of the process is entirely in the memory kernel, which accounts for how the information from the past affects the state of the system on aggregate. The local driving term, however, can be responsible for heightening the complexity. In fact, the master equation for two different drivings can be drastically different. This master equation also accounts for stochastic interventions from the past; these effects are included in the initial correlation (i.c.) term, as well as the memory kernel.

While there is some progress in understanding the computational complexity of the above master equation, it is largely in terms of the dilated complexity. However, there are recent works that bound the dimension of the environment from below by borrowing on similar methods from classical probability theory~\cite{khan2021model}. We note here in passing that all of our results focus on the dynamics of small systems; it is well-known that the open dynamics of many-body quantum systems can encapsulate all of quantum computing~\cite{verstraete2009quantum}.

Here, we will consider the computational complexity of the underlying quantum stochastic process to determine whether a master equation or a stochastic map is classically tractable. We first introduce the process tensor framework, which describes quantum stochastic processes in full generality, and then we will argue that the sampling complexity of the process is indicative of the complexity of the corresponding family of master equations.
The task of approximately sampling the output probability distribution of a quantum state has been considered a good representation for what it means for a classical computer to simulate a quantum system -- capturing the exact obtainable behaviour of a quantum experiment. This task is the foundation of the conventional \textit{quantum sampling problem}, and has been shown to be classically hard for a number of quantum systems~\cite{aaronson2011computational, bremner2011classical, bouland2019complexity}.

To reach our goal, we need to account for the interaction of a system with an environment that is often out of the control of the experimenter, as well as capture interesting dynamical properties of the process, such as temporal entanglement and causal relations. Just as sampling an entire many-body state allows us to infer all $k$-body spatial relations, we must sample our system across many times to capture all multi-time correlations arising from strong interaction of a system with a complex environment~\cite{pollock2018operational, pollock2018non, costa2016quantum, shrapnel2018updating, milz2020kolmogorov, milz2021quantum}. Our first task is, therefore, to generalise state-based sampling complexity to a temporal version capable of capturing the rich multi-time physics of open quantum systems. This is precisely what the process tensor enables -- namely, it has an equivalent representation as a quantum state. From this, we show that quantum sampling problems are not only a motivating tool for our newly defined temporal sampling problem, but can be formally placed on equal footing. 

\subsection{The process tensor formalism}
\begin{figure}[t]
    \centering
    \includegraphics[width=\linewidth]{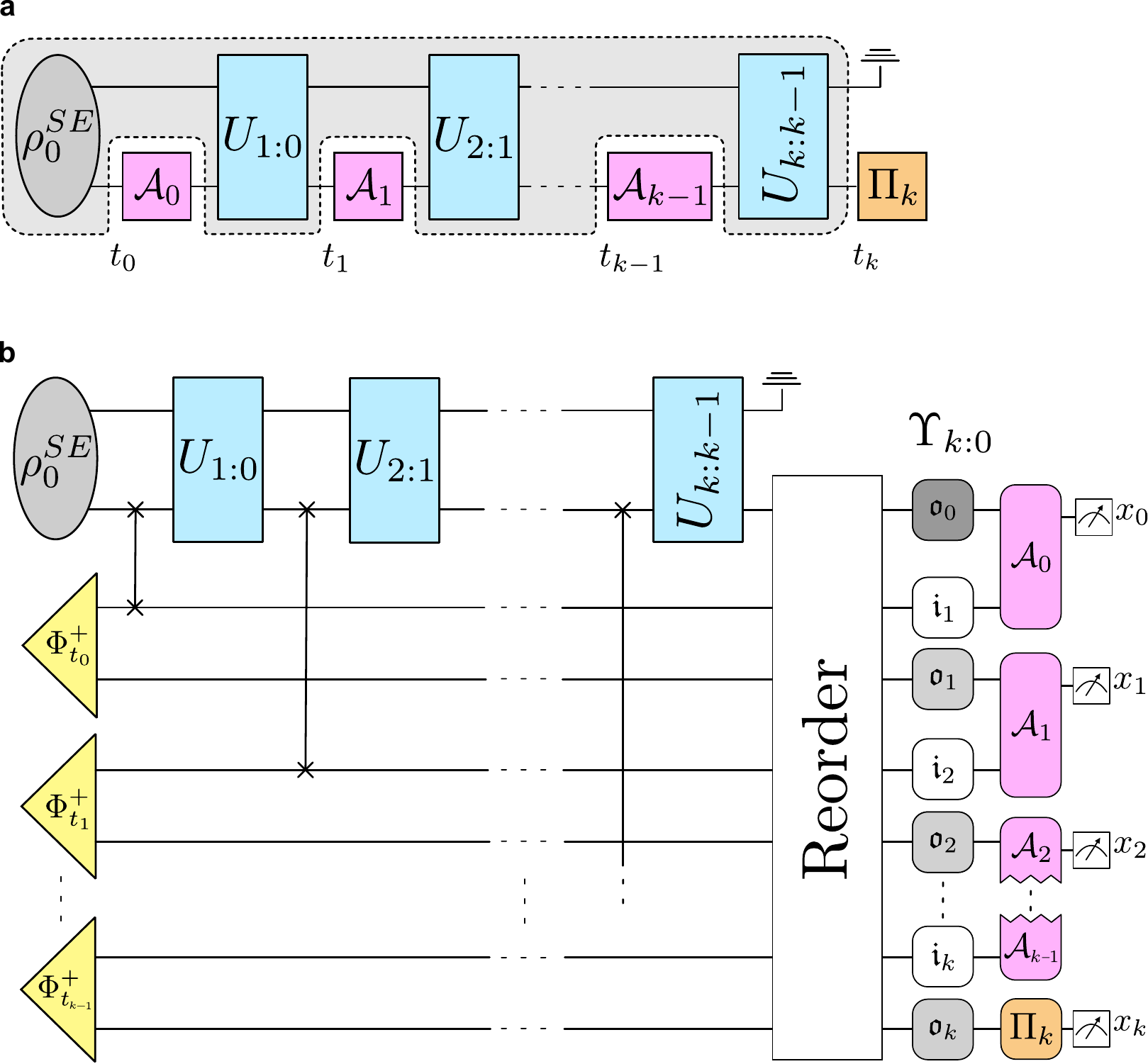}
    \caption{The process tensor description of open quantum systems. \textbf{a} A circuit form depiction of the process tensor. The unitaries $U_{j:j-1}$ describe arbitrary $SE$ dynamics. Control operations $\mathcal{A}_j$ are implemented at times $t_j$. Measurement outcomes at each time generate a probability distribution conditioned on the set of controlled operations. \textbf{b} A $k$-step process tensor Choi state. One half of fresh Bell pair $\Phi_{t_j}^{+}$ is swapped in at each time $t_j$. Temporal correlations are then mapped to spatial correlations in the Choi state. Projecting the Choi state onto the Choi state of the controlled operations corresponds to sampling from the many-body state.}
    \label{fig:Process tensor}
\end{figure}

Consider the setup depicted in Figure~\ref{fig:Process tensor}a. An initial system coupled to an environment evolves freely as described by unitary dynamics, $U_{j:j-1}$ between times $t_{j-1}$ and $t_{j}$, where $U_{j:j-1}\left(\cdot\right) = u_{j:j-1}\left(\cdot\right)u_{j:j-1}^{\dagger}$. Here, we consider the general case of a $d_s$-dimensional system interacting with a $d_e$-dimensional environment. At successive times, $t_{0} < t_{1} < \cdots < t_{k}$, the experimenter can probe the system with instruments to gain information about its properties. Each manipulation of the system at time $t_{j}$ corresponds to a trace non-increasing completely positive (CP) map, $\mathcal{A}_{j}$, representing any manner in which the experimenter chooses to probe the system. For example, the experimenter could choose to measure the system according to the positive operator-valued measure (POVM) $M_{x_j}$, and then on observing outcome $x_{j}$ feed forward the quantum state $\rho_{x_j}$. Here, the map describing the experimental manipulation of the system is given by $\mathcal{A}^{(x_j)}_{j}[\rho] = \text{Tr}(M_{x_j}\rho)\rho_{x_j}$.

More generally, the probability of observing outcomes $\mathbf{x}_{k:0}:=\{x_{k}, x_{k-1}, \ldots ,x_{0}\}$ following the application of instruments $\mathbf{J}_{k:0}:=\{\mathcal{J}_{k}, \mathcal{J}_{k-1},\ldots, \mathcal{J}_{0}\}$ represented by the set of CP maps $\mathbf{A}_{k-1:0} :=\{\mathcal{A}_{k-1},\ldots, \mathcal{A}_{0}\}$ and measurement apparatus $\{\Pi_k^{x_k}\}$ at times $\mathbf{T}_{k:0}:=\{t_{k},t_{k-1},\ldots,t_{0}\}$ is given by 
\begin{equation}
\label{eq:PT_probability}
\begin{aligned}
    \mathbb{P}\left(\mathbf{x}_{k:0}|\mathbf{J}_{k:0}\right) &=  \text{Tr}\left(\Pi_{k}^{(x_k)} \circ U_{k} \circ \cdots \circ U_{1} \circ \mathcal{A}_{0}^{(x_0)} \left[\rho_{0}^{SE} \right]\right)
    \\ &\coloneqq \text{Tr}\left(\mathcal{T}_{k}\left[\Pi_{k}^{(x_k)},\ldots,\mathcal{A}_{0}^{(x_0)}\right]\right), 
\end{aligned}
\end{equation}
where we have shortened our notation $U_{j:j-1} := U_j$, and introduced the multi-linear functional $\mathcal{T}_{k}$, known as the process tensor. Note that the instruments here only act on the system space, whereas each $U_j$ acts on both system and environment.

Equation~\eqref{eq:PT_probability} forms the basis of our definition for simulation complexity. 
Specifically, these constitute the probability distributions from which we will sample.
The process tensor, $\mathcal{T}_{k}$, represents the uncontrollable underlying evolution of an open quantum system and its environment, and is depicted by the light grey box in Figure~\ref{fig:Process tensor}a. The process tensor formalism is the natural generalisation of the theory classical stochastic processes~\cite{milz2020kolmogorov}. In other words, it contains all spatio-temporal correlations that the process can exhibit, be they classical or quantum. This has led to a variety of investigations into the nature of temporal correlations, e.g. when does a quantum process contain only classical correlations?~\cite{strasberg2019classical,milz2020non}; when do the correlations in a quantum process vanish?~\cite{figueroa2019almost,figueroa2021markovianization,dowling2021relaxation,dowling2021equilibration}; when are the correlations entangling?~\cite{costa2016quantum,milz2021genuine, ringbauer2018multi,giarmatzi2021witnessing}. In this work, we ask a complementary question, when are the correlations in a quantum process classically hard?

To do so, we consider sampling from a quantum process with operations $\mathcal{A}_j$. These allow us to infer all statistical information about the underlying process, as well as compute quantities such as local observables of the system at each time, $t_j$. Sampling from the distribution in Equation~\eqref{eq:PT_probability}, therefore, gives us all the properties of an open quantum system that we could hope to gain with an experiment, making it the ideal candidate for representing the simulation of these systems. We refer to this task as a dynamic sampling problem. Note that characterisation of distributions of this type has already been shown to be possible on quantum devices; our work has immediate implications for real-world analysis of open dynamics~\cite{white2020demonstration,white2021many,white2022non}.\par

Given a description of the initial $SE$ state, the set of unitaries $\textbf{U}_{k:1} := \left\{U_{k},\ldots,U_{1}\right\}$, and instruments $\textbf{J}_{k:0}$, we define the complexity of simulating an open quantum system as the classical complexity of sampling from the resulting temporal output distribution, $\mathcal{D}_{\textbf{J}}^{k}$, given by Equation~\eqref{eq:PT_probability}.  
This output distribution is conditioned on both the number of sampling times, $k+1$, and the choice of instruments, $\textbf{J}_k$, which form parameters of our complexity analysis. Consequently, the complexity of simulating an open quantum system may be classified as easy in some regimes, and hard in others. In fact, as we analyse different open quantum systems in Section~\ref{section:complexity of processes} and~\ref{section:Numerical results}, we identify simulation tasks that depend on the number of sampling times, and/or choice of instruments. The latter is of particular practical importance, since during any simulation of a process on real hardware, it is impossible to separate the operations required for implementing the $SE$ dynamics from the operations required to measure the system and gain useful information about the process. 

\subsection{The equivalence of multi-time and many-body sampling}
\label{sec:Choi representation}
Through a generalisation of the Choi-Jamiolkowski isomorphism (CJI), the process tensor mapping $\mathcal{T}_k$ is in one-to-one correspondence with a many-body state $\Upsilon_{k:0}$. Consequently, every dynamic sampling problem has an equivalent representation as a state-based sampling problem. 
The state representation of the process may be realised via the circuit shown in Figure~\ref{fig:Process tensor}b. At each $t_j$, one half of a fresh maximally entangled pair is passed through the corresponding $U_j$. Temporal correlations are then mapped onto spatial correlations between each of the subsystems. The resulting Choi state $\Upsilon_{k:0}$ constitutes a $2k+1$-partite state shown by the light grey box in Figure~\ref{fig:Process tensor}b. Similarly, each of the CP maps $\mathcal{A}_{j}$ correspond via the CJI to a matrix $\hat{\mathcal{A}}_{j}$. Equation~\eqref{eq:PT_probability} can then be rewritten as~\cite{milz2021genuine}:
\begin{equation}
    \label{eq: Spatiotemporal Born rule}
    \mathbb{P}\left(x_{k},\ldots,x_{0}|\mathcal{J}_{k},\ldots,\mathcal{J}_{0}\right) =  \text{Tr}\left[\Upsilon_{k:0}\left(\Pi_k \otimes  
    \hat{\mathbf{A}}_{k-1:0}^{\text{T}}\right)\right],
\end{equation}
where $\hat{\mathbf{A}}_{k-1:0} = \bigotimes_{j=0}^{k-1}\hat{\mathcal{A}}_j^{(x_j)}$

Equation~\eqref{eq: Spatiotemporal Born rule} represents the spatiotemporal generalisation of Born's rule~\cite{shrapnel2018updating, chiribella2011informational} and, consequently, demonstrates that operations on a system at successive points in time correspond to observables on the process tensor Choi state. The circuit in Figure~\ref{fig:Process tensor}b is, therefore, a sampling problem on a $2k + 1$ many-body state that is equivalent to a temporal sampling problem on a $k$-step process. 

\subsection{Master equations, stochastic maps, and process tensor}
We are now in position to re-examine master equations and stochastic maps, with an understanding for the basic structural elements of the process tensor. Firstly, it is clear that given a process tensor $\Upsilon_{k:0}$, we can obtain all stochastic maps $\{\Lambda_{j:i}\}$ with $0 \le i \le j \le k$. This is trivially obtained by choosing all $\mathcal{J}$ to be the identity instrument except at times $i$ and $j$. 
Next, it was recently shown that this same family of stochastic maps serves as a discrete version of the Nakajima-Zwanzig master equation by means of the transfer tensor method~\cite{cerrillo2014non, pollock2018tomographically, jorgensen2020discrete, milz2018reconstructing}. Thus, the process tensor contains both master equations and stochastic maps as limiting cases. One only has to make sure that the time resolution is fine enough to well-approximate the time derivative in Eq.~\eqref{eq:me}.

Importantly, if we consider the process tensor as a many-body state, then a master equation is sampling from that state by contracting it with the Choi states of instruments at all times except $i$ and $j$. The Choi state corresponding to the identity instrument is the state $\ket{\psi^+} := \sum_i \ket{ii}$; in fact, for local driving the Choi state is a maximally entangled state of the form $\ket{\psi^+(H_t)} := (\exp\{-i H_t\} \otimes \openone) \sum_i \ket{ii}$, meaning the instrument will change in time. Thus, any master equation is the same as sampling from the projected Choi state of the process tensor to the spaces of $i$ and $j$. If this sampling task is hard, then so is simulation of the master equation.

Thus, going forward we only consider the computational complexity of the process tensor. This is because the complexity of the process tensor represents the complexity of the underlying quantum process. If this is hard, then -- we posit -- so will be the family of master equations that are contained therein. Our strategy will be to show that, for a given process, there are sampling regimes that are classically hard. We do this for Shor's algorithm, the Heisenberg Hamiltonian, and several well-known models of quantum computation. We also then present numerical evidence for the hardness of process tensors generated by random circuits and random Hamiltonians.


\section{The complexity class \texorpdfstring{$\mathsf{OpenDQP}$}{OpenDQP}}
\label{section:OpenDQP}
With a clear definition for the task of simulating open quantum systems, we now turn to classifying their complexity. We are particularly interested in studying the boundary between when this task is classically easy versus hard, since we can consider this the point at which a system displays truly complex or quantum features. This corresponds to the boundary between the complexity classes $\mathsf{BPP}$ and $\mathsf{BQP}$, the class of problems solvable by a probabilistic classical and quantum computer, respectively. To formally relate examples of open quantum systems to these classes we define the complexity class \textit{Open Dynamic Quantum Polynomial-Time} ($\mathsf{OpenDQP}$), which encapsulates the set of problems solvable in our model.  

The dynamic sampling problem we have described bears close resemblance to the complexity class \textit{Dynamic Quantum Polynomial-Time} ($\mathsf{DQP}$) defined by Aaronson in~\cite{aaronson2005quantum}. $\mathsf{DQP}$ is the class of problems solvable by a $\mathsf{BQP}$ machine when given access to an oracle that can return a sample over the probability distribution of the classical histories of a quantum circuit. Using this model, Aaronson shows that $\mathsf{DQP}$ is slightly more powerful than $\mathsf{BQP}$. However, the ability to sample a circuit's histories without actually implementing a measurement is of course unphysical. 
Motivated by the study of physically implementing a dynamic sampling problem on a quantum computer, we define the analogous complexity class $\mathsf{OpenDQP}$ as follows.
\begin{definition}
$\mathsf{OpenDQP}$: The class of sampling problems S = $\{\mathcal{D}(\mathbf{x})_{\mathbf{J}}^{k}\}_{x \in \{0,1\}^{k}}$ for which there exists a quantum polynomial time algorithm Q, such that given the initial state $\rho_{0}^{SE}$ of size $n$, unitaries $\mathbf{U}_{k:0}$, and instruments $\mathbf{J}_{k:0}$ as input, outputs a sample $\textbf{\textup{x}} = \{x_{k},\ldots,x_{0}\}$ from distribution $\mathcal{R}(\textbf{\textup{x}})$, in time \textup{poly($n,k$)}, such that 
\begin{gather}
    \|\mathcal{R}(\textbf{\textup{x}}) - \mathcal{D}(\textbf{\textup{x}})_{\textbf{J}}^{k}\|_{1} \leq \epsilon 
    \quad \mbox{for} 
    \quad \epsilon = O\left(\frac{1}{\textup{poly(}k\textup{)}}\right).
\end{gather}
\end{definition}
It is important to note that there must exist an efficient classical description of the initial state and unitaries $U_{j}$, otherwise there could exist an environment that could compute problems outside of $\mathsf{BQP}$ and return the answer to the system. We, therefore, require unitaries of size \textup{poly($n$)}.
A process is then efficiently classically simulable if there exists a polynomial-time classical algorithm that outputs a sample from distribution $\mathcal{R}(\textbf{\textup{x}})$ -- i.e. one that can sample from the output distribution of the process up to additive error. This is consistent with the notion of ``weak'' classical simulation, which achieves polynomial accuracy in polynomial time (see~\cite{ni2013commuting}, particularly section 2.3, for further discussion). It follows that our sampling problem can be considered \textit{easy(hard)} if it can(not) be efficiently classically simulated. We will refer to a process as hard (or complex) if there exist a set of dependencies $\textbf{J}_k$, or number of sampling times such that it is classically hard to simulate. 

When considering the complexity of master equations, it is useful to define the complexity class $\mathsf{OpenDQP}$ in terms of the total number of sampling times $k$. We denote this $\mathsf{OpenDQP}_k$. 
Non-driven master equations and those with a constant local field then form the subset $\mathsf{OpenDQP}_2 \subseteq \mathsf{OpenDQP}_k$. 
These correspond to a $k$-step process tensor with identity instruments or identical fixed operations inserted at all times except $i$ and $j$ in order to construct the stochastic map $\Lambda_{j:i}$.
We include constant fields in this description, since these time-independent operations could be incorporated into the underlying system-environment interaction. 
By comparison, a driven master equation will have a set of time-dependent control operations $\mathcal{A}_j$, and thus exists within $\mathsf{OpenDQP}_k$.
These sets then form the family of master equations associated with an underlying quantum stochastic process.


\section{Complexity of Processes}
\label{section:complexity of processes}
Since the task of classically simulating any open quantum system can be framed as a problem within $\mathsf{OpenDQP}$, the computational power of this complexity class gives an indication of how hard it is to simulate these systems. In this section, we begin by establishing the relationship of $\mathsf{OpenDQP}$ to existing complexity classes to broadly classify its computational power. Subsequently, we consider a number of specific open quantum systems, in order to identify instances of $\mathsf{OpenDQP}$ that are classically hard to simulate. 



\subsection{\texorpdfstring{$\mathsf{BQP} \subseteq \mathsf{OpenDQP} = \mathsf{SampBQP}$}{BQP = OpenDQP = SampBQP} }
It immediately follows from our definition that $\mathsf{BQP} \subseteq \mathsf{OpenDQP}$, and thus $\mathsf{OpenDQP}$ is at least as powerful as standard quantum computation. $\mathsf{BQP}$ are the set of decision problems solvable by measuring a single qubit with a bounded error rate~\cite{lund2017quantum}. Thus, any problem in $\mathsf{BQP}$ can be represented as a sampling problem in $\mathsf{OpenDQP}$ restricted to a single time. If we prepare an initial state $|0\rangle^{\otimes n}$, apply a polynomial-size quantum circuit $U_{1:0}$, and then measure in the computational basis on the system at time $t_1$, we recover $\mathsf{BQP}$.

\begin{figure}[ht]
    \centering
    \includegraphics[width=\linewidth]{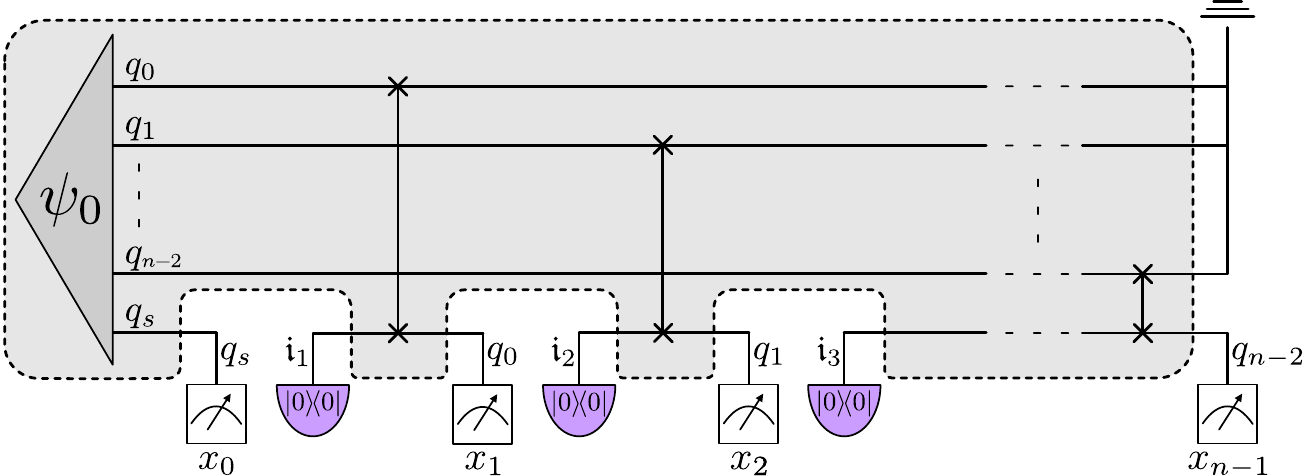}
    \caption{A process with temporal probability distribution equivalent to the spatial distribution of the initial state $\psi_{0}$. A measurement at time $t_j$, constitutes a measurement of an environment qubit (here indexed as $q_{j-1}$) and results in outcome $x_j$. A new state $|0\rangle\!\langle0|$ is fed into the subsequent input leg of the process tensor.
    }
    \label{fig:SWAP circuit}
\end{figure}

We then observe that $\mathsf{SampBQP} = \mathsf{OpenDQP}$; that is, we can represent any conventional state-based sampling problem as a dynamic sampling problem on a multi-time process, and vice-versa. Equivalently, this result states that quantum processes and quantum states can be equally complex. We showed that $\mathsf{OpenDQP} \subseteq \mathsf{SampBQP}$ in Section~\ref{section:OpenDQP}, and thus states are at least as complex as processes. Sampling from any multi-time process can be represented as sampling from a many-body state via the CJI, where operations at different times correspond to observables on the Choi state. To show the reverse inclusion -- that $\mathsf{SampBQP} \subseteq \mathsf{OpenDQP}$ -- we show that the output distribution of any state can be represented as the output distribution of a multi-time process. 

The logic follows the circuit diagram in Figure~\ref{fig:SWAP circuit}. Consider an arbitrary quantum sampling problem on a state. This is computed by taking an initial $SE$ $n$-qubit state $|0\rangle^{\otimes n}$ and evolving it via a circuit which corresponds to unitary $U$. At this point in time, the $\mathsf{SampBQP}$ problem involves sampling from the distribution of the output state given by $p(x) = |\langle x|U|0\rangle^{\otimes n}|^{2}$, where $x$ is a length $n$ bit-string. We label the resulting state $\psi_{0}$, and consider it the initial system-environment state now entering our process. We then measure the system resulting in outcome $x_0$. Next, we insert a SWAP gate acting between the first qubit of the environment, $q_0$, and the system, followed by a measurement and result $x_1$. This process is repeated for each of the $n-1$ qubits in the environment resulting in a $n$-step process tensor. Following the circuit lines, we can see that the probability of the $n$-length bit-string obtained from sampling the system, $(x_0,x_1,x_2,\ldots,x_{n-1})$ is precisely that of the state $\psi_{0}$.

Finally, we note that state complexity is necessary for process complexity. By this we specifically mean that the system-environment state must be hard to sample from at at least a single time during the dynamics. This hardness may be conditioned on a previous operation on the system. If this were not true, then we could directly simulate the $SE$ dynamics and compute the reduced output on the system.

\subsection{Illustrative example: Shor's algorithm \texorpdfstring{$\subseteq \mathsf{OpenDQP}$}{subset OpenDQP}}
Having established that $\mathsf{OpenDQP}$ is at least as powerful as quantum computation, we are particularly interested in finding problems that exist within $\mathsf{OpenDQP}$, but that are outside $\mathsf{BPP}$. This subset of problems represent quantum stochastic processes that are hard to classically simulate, and their existence provides evidence for a quantum advantage in simulating open quantum systems. Shor's algorithm for factoring integers is in the class $\mathsf{BQP}$, and provides an exponential speed up over the best known classical algorithms. As a result, factoring is widely believed to be outside $\mathsf{BPP}$, although there is not yet strong complexity-theoretic evidence for this~\cite{aaronson2011computational}. \par 

Here, we show that Shor's algorithm can, in fact, be cast as a sampling problem on a quantum stochastic process and provides a natural introduction to the conceptual idea of mapping state complexity onto process complexity through the process tensor. This, in turn, shows that there are quantum stochastic processes that are classically hard, provided factoring is classically hard. 
In fact, Kitaev's one controlling-qubit trick for factoring an integer already serves as a pre-existing example of the conversion from state sampling to a dynamic sampling problem~\cite{kitaev1995quantum, parker2000efficient, beauregard2002circuit}.

\begin{figure}[ht]
    \centering
    \includegraphics[width=\linewidth]{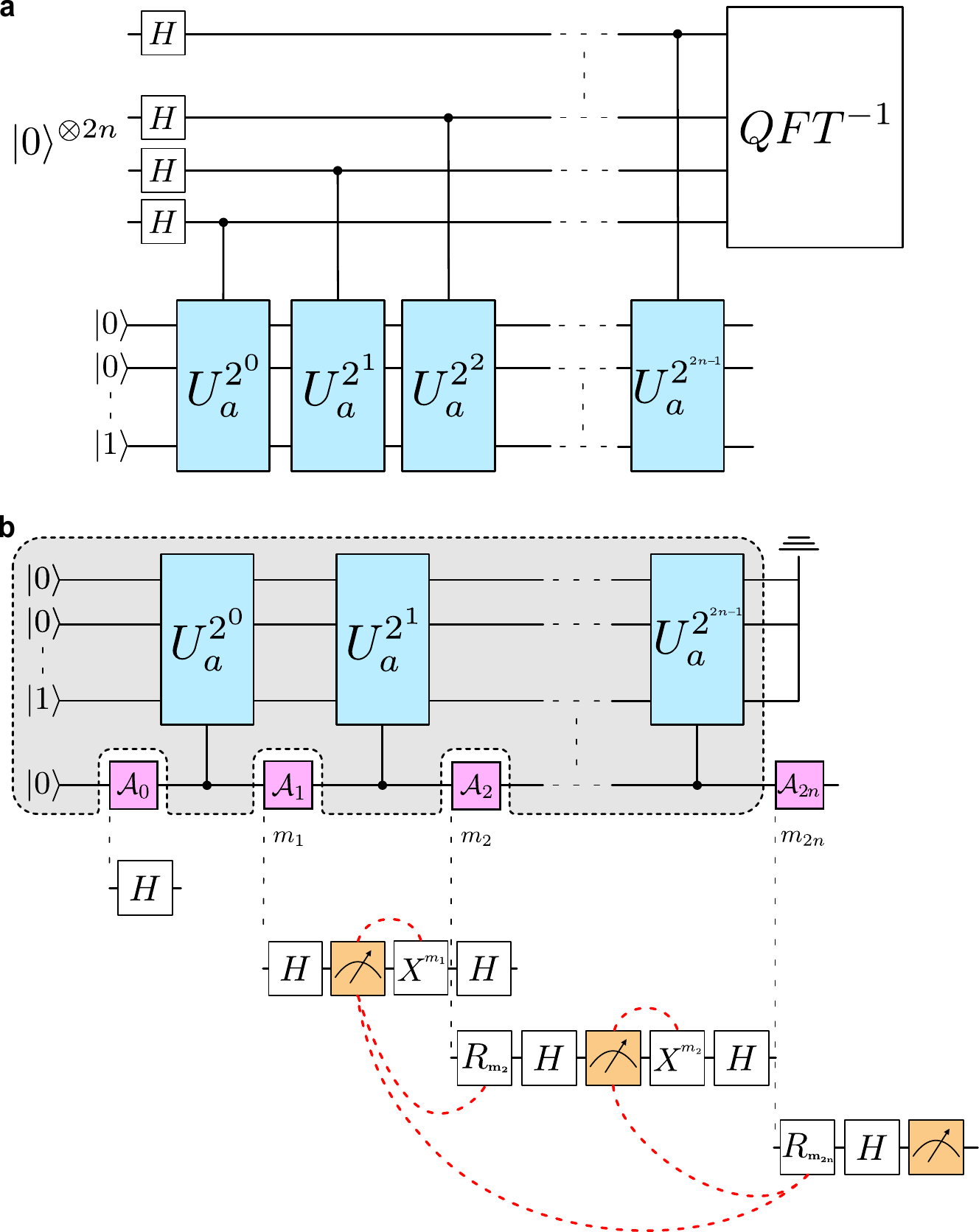}
    \caption{Shor's algorithm. 
    \textbf{a} The order finding circuit for Shor's factoring algorithm of a $n$-bit integer as described in the text. The unitaries perform modular multiplication and collectively transfor $|x\rangle$ to $|(ax) \bmod N\rangle$. The inverse QFT, subsequent measurement, and classical post processing gives the period of the function $f(x) = a^x \bmod N$.
    \textbf{b} Shor's Algorithm as a process tensor. The $2n$ register is collapsed to a single control or `system' qubit using the one controlled-qubit trick. The operations $\mathcal{A}_j$ correspond to the gates of the inverse QFT on a single qubit. At each time $t_j$, a rotation $R_{m_j}$ is performed which depends on all previous measurement results, followed by a measurement and reset  $X^{m_j}$, as described in the text. The classical correlations between these operations are shown by the red dotted line.}
    \label{fig:Shors}
\end{figure}

The order finding circuit for factoring an $n$-bit integer, $N$, is shown in Figure~\ref{fig:Shors}a. Here, the upper register of $2n$ qubits (where $n = \log N$) is a quantum superposition of integers $0,...,N^2 - 1$, followed by a modular exponentiation circuit of $2n$ controlled modular multiplications $U_{a}^{2^j}$, which collectively transform $\ket{x}$ to $\ket{(ax)\bmod N}$. The inverse quantum Fourier transform (QFT) circuit is then applied to the upper register, followed by a measurement and classical processing which gives the period of the function $f(x) = a^{x}\bmod N$. 

It has been shown that the factoring algorithm can be implemented in a semi-classical manner on a single qubit, plus an ancilla register~\cite{parker2000efficient, mosca1998hidden}. We will summarise the procedure briefly here as it is important for understanding Shor's algorithm as a process tensor. The key is to condense the entire $\ket{x}$ register into a single qubit, reset and reused. At each $t_j$, the system qubit acts as the $x_j$ bit of the register, placed into a $\ket{+}$ state and controlling the unitary $U_a^{2^j}$ on the ancilla space. Collectively, this computes $a^x \text{ mod }N$. The inverse QFT is then performed semi-classically: the control qubit is measured in the Fourier basis and re-prepared into a $\ket{+}$ state. Measurement in the Fourier basis involves a series of $Z$ rotations conditioned on outcomes at the previous times, identical to the controlled-$Z$ rotations in the standard inverse QFT circuit. This procedure is summarised in Figure~\ref{fig:Shors}b. Importantly, the specific choice of instruments and resulting bit-string can be used to efficiently determine the prime factors of a number. This leads us to our first result:

\begin{result}
Shor's algorithm to factor an $n$-bit number can be written as a sampling problem from an open quantum system. This quantum stochastic process is represented by a $2n$-step single qubit process tensor.
\label{result:Shor}
\end{result}

To see this, we first note that the order finding circuit of Figure~\ref{fig:Shors}a can be represented as a $k=2n$ step process in its generalised Choi state form.  The $n$-qubit register is now the environment, and the $2n$ qubits of the previous upper register each form one half of a maximally entangled Bell pair $|\Phi^{+}_{t_j}\rangle = \frac{1}{\sqrt{2}}(|00\rangle + |11\rangle)$, which is fed into the modular exponentiation circuit at time $t_j$. The resulting $2k+1$-body state is the process tensor Choi state $\Upsilon_{k:0}$ of Shor's algorithm. The inverse QFT coincide with operations on this Choi state or, equivalently, sampling of the many-body state. The mathematical construction of the state $\Upsilon_{k:0}$ is detailed in Appendix~\ref{app:Shor's choi state}.  

We can then utilise Kitaev's proposal to map this state sampling problem, to a dynamic sampling problem on a multi-time process. This is represented in Figure~\ref{fig:Shors}b. The process tensor, shown in light grey, corresponds to unitary evolution of a system-environment according to the controlled modular multiplication circuits. After each controlled unitary, we can probe our system with the semi-classical operations corresponding to the inverse QFT on a single qubit. At time $t_j$, if we observe outcome $x_j = \{0,1\}$, then the map representing this transformation is given by $\mathcal{A}_j^{x_j}[\rho] = \text{Tr}\left(M_{x_j}\left(R\rho R^{\dagger}\right)\right) |+\rangle\!\langle+|$, which accounts for a rotation gate conditioned on all previous classical measurement outcomes, a POVM $M_{x_j}$ corresponding to measuring in the $X$ basis and observing outcome $x_j$, and feed forward of the state $|+\rangle$ for use at the next time-step. Shor's algorithm is, therefore, a problem that is clearly contained in $\mathsf{OpenDQP}$ and provides a tangible example that suggests within this framework there will exist open quantum dynamics that are classically hard to simulate in the worst case.

A nice consequence of framing Shor's algorithm as a quantum stochastic process, is that we identify an instance of $\mathsf{OpenDQP}$, where the classical complexity depends on the choice of sampling operations, $\mathcal{A}_j$. This establishes our second result:

\begin{result}
\label{result: Shor sampling}
The process tensor, from which Shor's algorithm is derived, is easy to sample from in the computational basis.
\end{result}

The proof of this follows from observing the circuit shown in Figure~\ref{fig:Shors}b. If instead of performing the series of measurements, rotation gates, and Hadamard, which implement the inverse QFT, we simply measure in the computational basis at each time, then the circuit on the system alone corresponds to sampling on a single qubit at each time following the implementation of a Hadamard gate. Thus, the state of the system prior to any measurement is $|+\rangle = \frac{1}{\sqrt{2}} (|0\rangle + |1\rangle)$, and classical simulation of the output of the process tensor is given by the outcome of a coin toss at each time. 

Ultimately, there exists a process tensor version of Shor's algorithm which is easy to sample from in the computational basis, but can factor numbers if we sample with the classically correlated control operations described in Figure~\ref{fig:Shors}b. It is quite remarkable that manipulations on a single qubit alone can elevate the complexity of simulation from a simple classical stochastic process solvable in polynomial-time, to one solvable only in quantum polynomial-time.

\subsection{Physical Hamiltonian: Heisenberg interaction \texorpdfstring{$\subseteq \mathsf{OpenDQP} \setminus \mathsf{BPP}$}{subset OpenDQP}} \label{sec:Heisenberg}
Shor's algorithm provides an introductory example of a sampling problem in $\mathsf{OpenDQP}$ that is classically hard to simulate given a specific choice of operations. 
However, we are primarily interested in the simulation of physical systems. 
Here we present a physical Hamiltonian for the evolution of a system and environment, which along with a set of sampling operations $\mathcal{A}_j$ on the system, constitutes a sampling problem in $\mathsf{OpenDQP}$ that is in $\mathsf{BQP} \setminus \mathsf{BPP}$ and is thus classically hard to simulate. 
Specifically, we show that the Heisenberg interaction in a spin chain coupled with operations on the system results in a temporal probability distribution that is hard to sample from. 
The general idea is that in this interaction picture we can use operations on the system to construct a complex system-environment state, and then extract this complexity dynamically.  

To see this, we first observe that there exist composite systems which admit algebraic control (AC) and can be framed as problems in the complexity class $\mathsf{OpenDQP}$.
The algebraic control approach to quantum control determines the necessary and sufficient conditions such that a composite system can be completely driven by control operations on only a smaller subsystem~\cite{romano2006incoherent}.
In this method there exists a composite system $V = S \cup E$, with controllable subsystem $S$ described by the global Hamiltonian $H_V + \sum_{j} f_j (t)h_{S}^{(j)} \otimes \openone_{E}$. 
The time-dependent local controls $h_S^{(j)}$ are achieved through the modulating parameters $f_j (t)$.

\begin{figure}[ht]
    \centering
    \includegraphics[width=\linewidth]{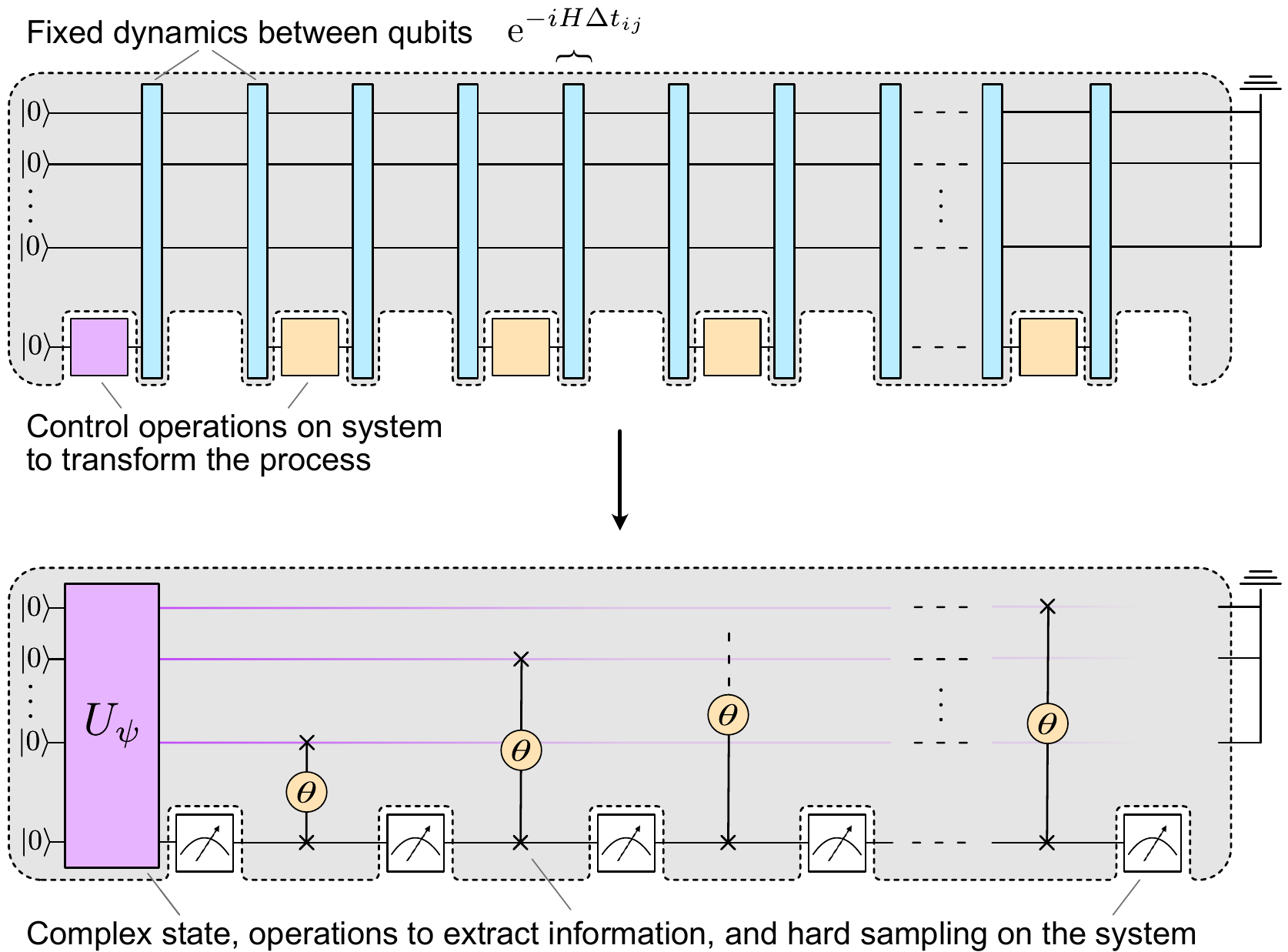}
    \caption{A complex dynamic sampling problem generated by the Heisenberg interaction on a spin chain. A single system qubit interacts with a spin chain with fixed Heisenberg coupling, $H$. The top figure shows a sequence of control operations on the system in pink and yellow, which can perform arbitrary transformations of the composite system-environment state. In the bottom figure these operations result in evolution of the $SE$ state given by $U_\psi$, which results in the spatially complex state $\psi$. This is followed by a series of pSWAP gates between the system and environment. The pSWAP gates transfer information from the complex environment to the system, and the corresponding measurements on the system constitute a classically hard sampling problem.}
    \label{fig:Heisenberg spin chain}
\end{figure}

It has been shown that Heisenberg-type chains of $N$ spin-1/2 particles with arbitrary coupling strengths admit algebraic control through operations on a single end of the chain~\cite{burgarth2009local, PhysRevX.7.041015}.
This can be represented precisely as a dynamic sampling problem in $\mathsf{OpenDQP}$.
Here, the combined system and environment form the composite system, with global Hamiltonian given by the Heisenberg coupling:
\begin{equation}
\label{eq:Heisenberg}
    H_{ij} = J_{ij} \left(X_i X_j + Y_i Y_j + Z_i Z_j\right),
\end{equation}
where the subscripts $ij$ denote neighbouring qubits, and $J_{ij}$ are the couplings. 
The local controls act only on the system, and correspond to time-dependent sampling operations.
The results of AC mean that in the Heisenberg interaction picture, operations on the system can generate the entire Lie algebra SU($2^N$) on the system-environment state, and are thus universal for quantum computing. 
The net outcome is that we can generate any quantum transformation on the spin chain through operations on the system alone.
In particular, we can generate a system-environment state which is classically hard to sample from.

We now want to show that this composite system under algebraic control permits a dynamic sampling problem in $\mathsf{OpenDQP}$ that is classically hard, thus demonstrating a physical Hamiltonian of an open quantum system that is hard to simulate.
For this problem, it is important that we can efficiently generate complex states using the algebraic control method. That is, if a state can be prepared in polynomial-time by a quantum computer through operations on the entirety of the system, then achieving the same control through operations on the subsystem must also scale polynomially. 
Unfortunately, there is no general solution determining if this quantum control can be achieved efficiently. 
For the general Heisenberg interaction, solving for the controllable operations in itself is inefficient because the Hamiltonian cannot be diagonalised straightforwardly. 
Thus, commenting on the efficiency of the resulting operations is a highly non-trivial problem.
However, progress has been made presenting scalable implementation with control on two qubits and a tunable magnetic field on the environment~\cite{burgarth2010scalable}.
Moreover, local operator control on only a single qubit in a Heisenberg spin chain has been demonstrated to high fidelity on a chain of length 3~\cite{heule2010local}.
Thus, although implementing any unitary will take an exponentially long time, we instead only required the implementation of a unitary which takes our state to one outside of $\mathsf{BPP}$.
We anticipate that there will exist complex states that can be efficiently generated by operations on the system alone, and assume this to be true in what follows. 

We construct our sampling problem as per the circuit structure shown in Figure~\ref{fig:Heisenberg spin chain}.
We begin with an initial system-environment with fixed nearest neighbour Heisenberg coupling.
Using the algebraic controllability of this composite system-environment, we can then implement a series of operations $\mathcal{A}_j$ on the system at times $t_j$ until we have transformed the system-environment state into a complex state that is classically hard to sample from, i.e. one that is in $\mathsf{SampBQP} \setminus \mathsf{BPP}$. We label this state $\psi$. 

Using the same Heisenberg interaction we then leverage the spatial complexity of the state $\psi$ and redistribute the amplitudes of the environment onto the temporally sampled populations. Specifically, we use our control operations to engineer a series of partial-SWAP gates (pSWAP) between each environment qubit and the system, defined as: 
\begin{gather}
\mbox{pSWAP}(\theta)_{ij} := \exp\left\{-i\tfrac{\theta}{2}(X_i X_j + Y_i Y_j + Z_i Z_j)\right\},
\end{gather}
where the subscript $ij$ denotes qubits $i$ and $j$. These gates pass information directly from the environment to the system with probability depending on $\theta$. We consider a total of $k$ successive pSWAP gates between each environment qubit and the system, with sampling in the computational basis following each gate. This is repeated for all $n-1$ qubits of the environment, resulting in $(n-1)k$ sampling times.

The series of pSWAP gates between the environment and system results in a distribution on the system isomorphic to that of the constructed system-environment state $\psi$, and thus sampling from the system over time is as hard as sampling from this state. For simplicity, we first show that this is true for an environment of only one qubit. This can then be repeated iteratively for each qubit that makes up the environment, since we only interact with a single qubit at a time in our procedure. We consider the Choi state construction of a process with environment in some initial state $|\psi\rangle_{E} = \alpha|0\rangle + \beta|1\rangle$, where instead of feeding one half of a Bell pair in at each time, we feed the $|0\rangle$ state. This is equivalent to the construction using Bell pairs, and can be realised with the process tensor Choi state by projecting all input legs  onto $|0\rangle\!\langle0|$. At time $t_1$ we feed in the first ancilla $a_1 = |0\rangle_{1}$ and perform a pSWAP$_{E,a_{1}}$. Following this the environment-ancilla state is $|\psi\rangle_{Ea_1} = \alpha|00\rangle - ie^{i\theta}\beta\sin\theta|01\rangle + |10\rangle$, and the probability that the environment is in the zero state is $1 - |\beta|^2 \cos^{2}\theta$. At time $t_2$, the next ancilla $a_2 = |0\rangle_2$ is fed into the process, followed by another pSWAP. The feeding in of ancillas is equivalent to measuring the system and freshly resetting it to the $\ket{0}$ state.
This is repeated $k$ times, at which point the probability that the environment is in the zero state is given by $\text{Pr}[|\psi\rangle_E = |0\rangle] = 1 - |\beta|^2 \cos^{2k}\theta$. For a given choice of $\theta$, if $k = O(\log \epsilon)$, where $\epsilon > 0$, then the probability that the environment is in the zero state is:
\begin{equation}
    \text{Pr}\left[|\psi\rangle_E = |0\rangle\right] > 1 - \epsilon.
\end{equation}
Ultimately, the environment is in the $|0\rangle$ state with probability $1$ up to additive error, and all operations performed are unitary, thus the resulting distribution on the process tensor Choi state is isomorphic to the system-environment state $\psi$. 

As a result of this construction, we have identified an instance of $\mathsf{OpenDQP}$ that is equivalently hard to the sampling problem on the state $\psi$. Thus for any spatially complex state, this dynamic sampling problem is classically hard to simulate. This simple circuit is an illustrative example of the type of interactions which generate complex dynamics. Here, information from a complex environment is passed sequentially to the system, which is then measured. This behaviour resembles the physical scenario where complex memory effects in the environment can be temporally redistributed, affecting the future dynamics of the system. The temporal complexity here reflects how strongly the environmental complexity is imprinted upon the system.


\subsection{Clifford processes \texorpdfstring{$\subseteq \mathsf{OpenDQP} \setminus \mathsf{BPP}$}{Clifford in OpenDQP/BPP}}\label{sec:Clifford} 
\begin{figure}[t]
    \centering
    \includegraphics[width=\linewidth]{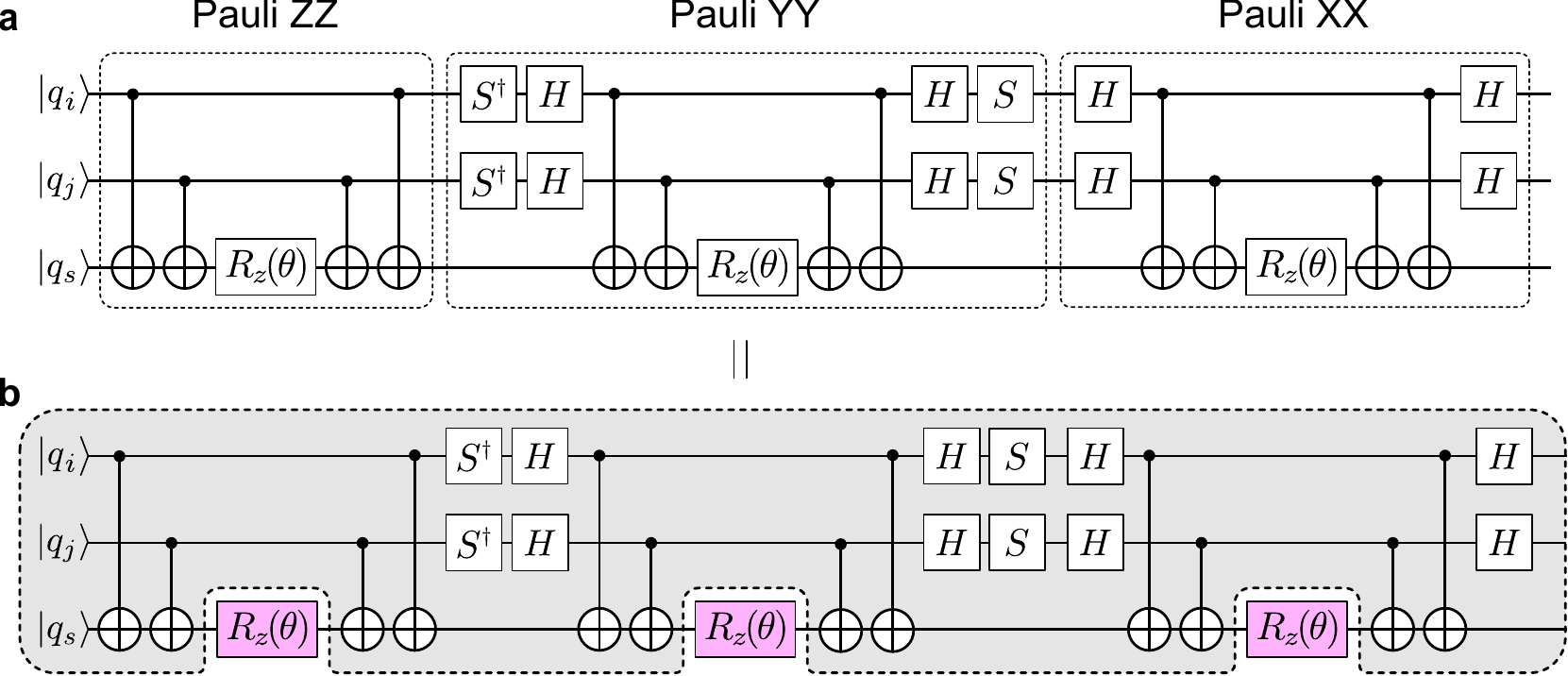}
    \caption{\textbf{a} Decomposition of a pSWAP gate into Clifford and $R_z$ gates. \textbf{b} The decomposition results in a process composed of Clifford gates, with sampling operations $R_z (\theta)$ acting on the system. For $\theta = \pi/4$ the sampling operations are $T$ gates.}
    \label{fig:partial SWAP decomposition}
\end{figure}

We can build on this example and decompose the Heisenberg couplings and pSWAP gates into Cliffords and non-Cliffords, which is shown for a single pSWAP in Figure~\ref{fig:partial SWAP decomposition}a. Treating the Cliffords as the background process here, the non-Cliffords can be designated as user-chosen instruments. As a control problem, this identifies a case in which the process tensor Choi state can be exactly constructed, but from which sampling is hard.
This establishes a sampling problem that is in $\mathsf{OpenDQP}$ but outside of $\mathsf{BPP}$. In other words, the degree of complexity of the sampling problem depends on the choice of operations, and can be controlled at the system level. 

First, we consider the pSWAP gates. A single pSWAP acting between two qubits $q_i$ and $q_j$ can be decomposed as follows. Since $XX$, $YY$, and $ZZ$ all commute, we can write the pSWAP as:
\begin{gather}
\exp(-i\tfrac{\theta}{2}X_i X_j) \exp(-i\tfrac{\theta}{2}Y_i Y_j)
\exp(-i\tfrac{\theta}{2}Z_i Z_j).
\end{gather}
Additionally, the $XX$ and $YY$ terms are locally equivalent to $ZZ$. A single pSWAP gate can then be represented as shown in Figure~\ref{fig:partial SWAP decomposition}a. We have now decomposed each pSWAP gate into a series of Clifford gates acting on the environment and system, plus three $T$ gates acting on the system only. These can be separated into the underlying $SE$ interaction, and sampling operations $\mathcal{A}_j$, as shown in Figure~\ref{fig:partial SWAP decomposition}b by the light grey box and pink $R_z$ gates, respectively. This procedure can be extended to all pSWAP gates, and similarly to the Heisenberg interaction of Equation~\eqref{eq:Heisenberg}. The overall result is a process tensor composed entirely of Clifford gates, and a sequence of sampling operations $\mathcal{A}_j$ made up of algebraic control operations, $T$ gates, and computational basis measurements.  

We identify the underlying process as an instance of $\mathsf{OpenDQP}$, where the $SE$ unitaries are Clifford circuits. By the Gottesman-Knill theorem~\cite{gottesman1998heisenberg}, if at each time we choose to measure our system in the computational basis, then sampling from the output of this Clifford process tensor is classically easy. However, as we have identified above, sampling with a sequence of algebraic control operations, $T$ gates, and computational basis measurements is equivalent to sampling from a complex state $\psi$, and thus classically hard. This again identifies an instance of $\mathsf{OpenDQP}$ where the choice of sampling operations can elevate our problem from one inside $\mathsf{BPP}$, to a sampling problem in $\mathsf{BQP} \setminus \mathsf{BPP}$. 

The ability to control complexity at the system level is particularly interesting when considering applications on near-term smaller scale quantum devices. Currently, we only have access to a small number of qubits, that are too noisy to generate the highly entangled system-environment states required for the simulation of complex dynamics. However, the noise in these devices is partly due to the interaction of each qubit with the surrounding environment, which in itself can generate complex statistics. Being able to control this interaction from a single qubit would allow the experimenter to access this complex noise as a resource, which may aid in the simulation of open quantum systems. In fact, the ability to modify complex temporal correlations has been experimentally verified on IBM devices, where changing control operations on a system qubit interacting with single qubit environment allows the experimenter to tune the resulting temporal entanglement~\cite{white2021many}.

\subsection{\texorpdfstring{$\mathsf{OpenDQP}\mbox{-}(\mathsf{IQP}) \not\subset$}{IQP not in} \texorpdfstring{$\mathsf{OpenDQP} \setminus \mathsf{BPP}$}{OpenDQP}}
Having found a member of $\mathsf{OpenDQP}$ that is outside of $\mathsf{BPP}$, and identified how a spatially complex process can be linked to a temporally complex process, we now investigate if all instances of $\mathsf{SampBQP}$ that demonstrate quantum advantage are progenitors of instances of $\mathsf{OpenDQP} \setminus \mathsf{BPP}$. By considering the \textit{``Instantaneous" Quantum Polynomial-time} ($\mathsf{IQP}$) framework, we show that this is not the case. $\mathsf{IQP}$ circuits, introduced and described in detail in~\cite{shepherd2009temporally, bremner2011classical}, are a restricted form of quantum computing comprising 2-qubit commuting gates which are diagonal in the $X$ basis. For simplicity, we restrict our attention to an equivalent description of $\mathsf{IQP}$ with gates diagonal in the computational basis. The circuit then consists of an initial $n$-qubit state $\ket{0}^{\otimes n}$, followed by $H^{\otimes n} C H^{\otimes n}$, where $C$ is a quantum circuit comprising gates diagonal in the $Z$ basis. Sampling an $\mathsf{IQP}$ circuit amounts to measuring in the computational basis on designated set of output lines. Bremner, Jozsa and Shepherd~\cite{bremner2011classical} showed that sampling on all $n$ output lines of an $\mathsf{IQP}$ circuit is hard to simulate classically unless the polynomial hierarchy collapses to the third level. However, measuring only $O(\log{}n)$ qubits is in $\mathsf{P}$.

\begin{figure}[ht]
    \centering
    \includegraphics[width=\linewidth]{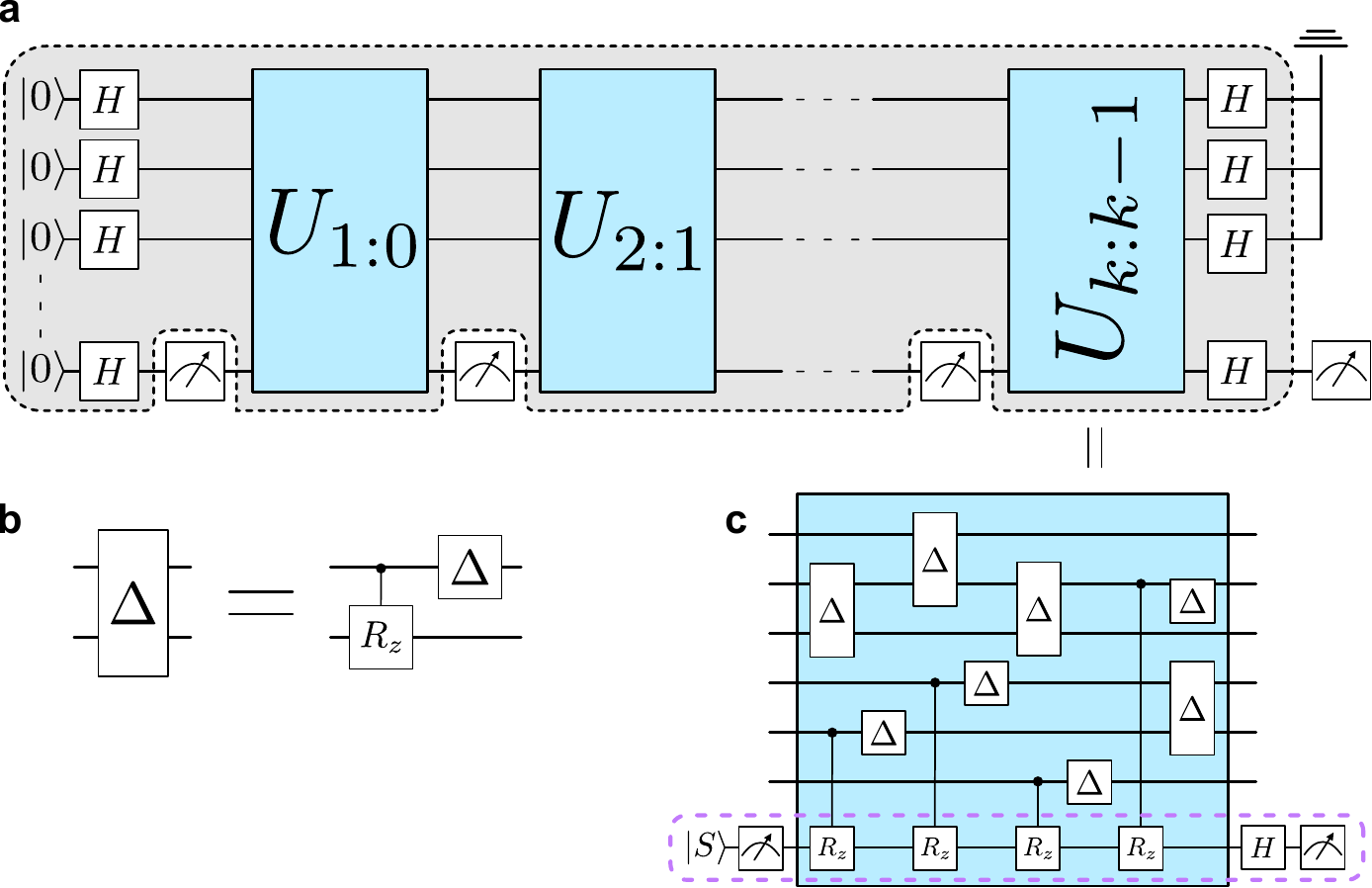}
    \caption{\textbf{a} The circuit construction of $\mathsf{OpenDQP}_{\mathsf{IQP}}$. \textbf{b} The decomposition of a 2-qubit $Z$-diagonal gate ($\Delta$) into a controlled-$R_z$ and single qubit $Z$-diagonal gate. \textbf{c} Each gate acting between an environment qubit and the system can be decomposed according to the description in b. The net result is a series of gates acting between environment qubits, and a series of $R_z$ gates acting on the system conditioned on environment qubits. The resulting operations acting on the system are outlined with the purple traced line.}
    \label{fig:IQP process}
\end{figure}

We now turn to our complexity analysis of $\mathsf{OpenDQP}$, and consider an instance where the underlying system-environment interaction (the unitaries $U_{j:j-1}$ in Figure \ref{fig:Process tensor}) is an $\mathsf{IQP}$ circuit on $n$ qubits, described in Figure \ref{fig:IQP process}. Here, we take a single qubit to represent our system, and the remaining $n-1$ qubits constitute the environment. We label this process $\mathsf{OpenDQP}\mbox{-}(\mathsf{IQP})$. The $\mathsf{IQP}$ circuit can be equivalently generated by a Hamiltonian diagonal in the $X$ basis which evolves the input state $\ket{0}^{\otimes n}$ according to~\cite{shepherd2009temporally}:
\begin{equation} \label{eq1}
H = \sum_{\textbf{p}} \bigotimes_{j:p_{j} = 1} h_jX_{j} 
\end{equation}
Here, there are a polynomial number of $n$ bit-strings \textbf{p}, which determine which qubits to apply the product of Pauli $X$ operators to in each term of the summation.

In this set-up, the results of Bremner, Jozsa and Shepherd confirm that the dynamics of the entire $SE$ state are hard to classically simulate. That is, the dynamics demonstrate spatial complexity. We now query whether the dynamics demonstrate temporal complexity. In particular, whether the temporal output distribution generated by the open quantum systems version of $\mathsf{IQP}$ is hard to sample from. We know that spatial complexity of the system-environment is a necessary condition in order to demonstrate temporal complexity on the system alone. 
We find that the process version of $\mathsf{IQP}$ can be efficiently classically simulated using Monte Carlo methods, resulting in an overall weak classical simulation of the process tensor for any number of times. This leads us to our third result:

\begin{result}
Temporally complex processes require a spatially complex $SE$ state; spatially complex $SE$ states are not sufficient to guarantee temporally complex processes.
\end{result}

To prove this we provide an efficient classical algorithm, following the methods in~\cite{bremner2011classical}, which outputs a sample from the temporal distribution generated by sampling the process $\mathsf{OpenDQP}\mbox{-}(\mathsf{IQP})$. Before proceeding with the algorithm, we provide some intuition for the types of processes which permit a Monte Carlo simulation by considering their representation as a Choi state. 
First, we note that Markovian processes have a product Choi state, and can be efficiently simulated~\cite{pollock2018operational, jozsa2003role}. A separable non-Markovian process, then, is a probabilistic mixture of Markovian processes. If the classical distribution governing the mixture can be efficiently sampled from, then it follows that the non-Markovian process can also be efficiently sampled from.
This is true for the process $\mathsf{OpenDQP}\mbox{-}(\mathsf{IQP})$.  

The proof is given here, with corresponding algorithm in Appendix~\ref{app:pseudocode IQP}. Consider the circuit in Figure~\ref{fig:IQP process}a on $n$ qubits. The lower qubit represents the system, and the upper qubits represents the $n-1$ sized environment. This is a $k$ step process with $k+1$ possible control operations. The $SE$ unitaries comprise $\text{poly}(n)$ $Z$ diagonal gates, which can be divided into two separate groups of 2-qubit diagonal gates: those acting between environment qubits, and those acting between the system qubit and an environment qubit. Each of these groups also consist of at most $\text{poly}(n)$ gates. Each gate in the latter group can be efficiently decomposed into a controlled-$R_{z}$ gate with the phase acting on the system qubit, plus an $R_{z}$ gate on the environment~\cite{shende2005synthesis,krol2021efficient}, as shown in Figure~\ref{fig:IQP process}b. To compute the output of this process tensor given a measurement in the computational basis at each time we proceed as follows: 

\begin{enumerate}
    \item Compile the $Z$-diagonal gates acting between the system and an environment qubit into poly(n) controlled $R_{z}$ gates and single qubit diagonal gates, where the $R_z$ unitary acts on the system qubit. 
    \item Consider the total system-environment state at the end of the circuit. Since this is an equal superposition state, tracing over the environment means we can stochastically replace it with a bit-string of length $n-1$ uniformly at random.
    \item This $n-1$ length bit-string commutes with the controlled $R_{z}$ system-environment gates at each of the time-steps, and thus the sample may be brought to the beginning of the circuit. We use this to determine the phase gates applied to the system qubit.
    \item The evolution of the single qubit system will then involve $\text{poly}(n)$ single qubit rotation gates $R_{z}$, plus $k+1$ total operations $\mathcal{A}_j$ at each time, which can be classically strongly simulated for each given weak sample of the process.
\end{enumerate}

We can now write the Choi state of this process as a mixture of separable states. For a $k$-step process this is given by:
\begin{equation}
\Upsilon_{k:0} =\frac{1}{2^{n-1}} \sum_{j=0}^{2^{n-1} - 1} \bigotimes_{i=1}^{k}  \ket{\Psi(\theta_{j}^{i})}\!\bra{\Psi(\theta_{j}^{i})} \otimes \rho_{0},
\end{equation}
where $\rho_{0}$ is the initial state of the system, and $j$ are the $2^{n-1}$ different $n-1$ bit binary strings representing the state of the environment. For a fixed bit $j$ at the $i$th time-step, $\theta_{j}^{i}$ represents the cumulative phase applied to one half of the maximally entangled Bell pair depending on the string $j$ which determines the control for the controlled-$R_{z}$ gates.  We then sum over all $n-1$ bit binary strings, which corresponds to tracing over the environment. 

From a physical perspective, the fact that $\mathsf{OpenDQP}\mbox{-}(\mathsf{IQP})$ is easy to sample from for any number of times and any choice of operations is surprising. We are probing a small part of a very complex system-environment interaction across multiple times that is, in itself, hard to sample from. Although there exists an equivalence relation between the complexity of the set of states and processes, there are clearly additional requirements on the system-environment interaction in order to leverage spatial complexity into temporal complexity. From this, we learn, at minimum, that a complex process must either generate temporal entanglement, or (if separable) be described as a mixture of trajectories where the probability distribution is hard to sample from.

It is worth noting that even though sampling from $\mathsf{OpenDQP}\mbox{-}(\mathsf{IQP})$ is an easy task, the related problem of learning the output distribution may still be classically hard. The difficulty with which quantum stochastic processes may be learned constitutes an interesting avenue for future work~\cite{shrapnel2018quantum, guo2020tensor, luchnikov2019machine, luchnikov2020machine}.

\section{Numerical evidence for Random Processes \texorpdfstring{$\subseteq \mathsf{OpenDQP} \setminus \mathsf{BPP}$}{OpenDQP/BPP}}
\label{section:Numerical results}

Having determined general characteristics of system-environment interactions that contribute to the classical difficulty of dynamic sampling we now return to the study of physical models. 
In section~\ref{sec:Heisenberg} we identified a physical Hamiltonian, the Heisenberg interaction, that in principal could result in a dynamic sampling problem that is classically hard.
We now consider two models that can be directly investigated for their complexity using numerical benchmarks, thus providing further verification that classical complexity is obtainable when sampling from physical models of open quantum evolution. 
They are random circuits sampling (RCS) and random matrix sampling.
The latter enjoys a wide range of physical applications~\cite{guhr1998random}, such as modeling quantum chaos~\cite{blumel1990random}, the statistical properties of atomic nuclei~\cite{french1988statistical}, and conductance in disordered mesoscopic systems~\cite{beenakker1997random}.
While, the former has been proven to demonstrate average-case hardness~\cite{bouland2019complexity} and is conjectured to be robust to realistic levels of noise~\cite{bouland2022noise}, which is essential for near-term implementations of quantum advantage on noisy quantum devices. This led to one of the first claimed demonstrations of quantum computational advantage through RCS on a 53-qubit superconducting quantum computer~\cite{arute2019quantum}.

\begin{figure*}[ht]
    \centering
    \includegraphics[width = \linewidth]{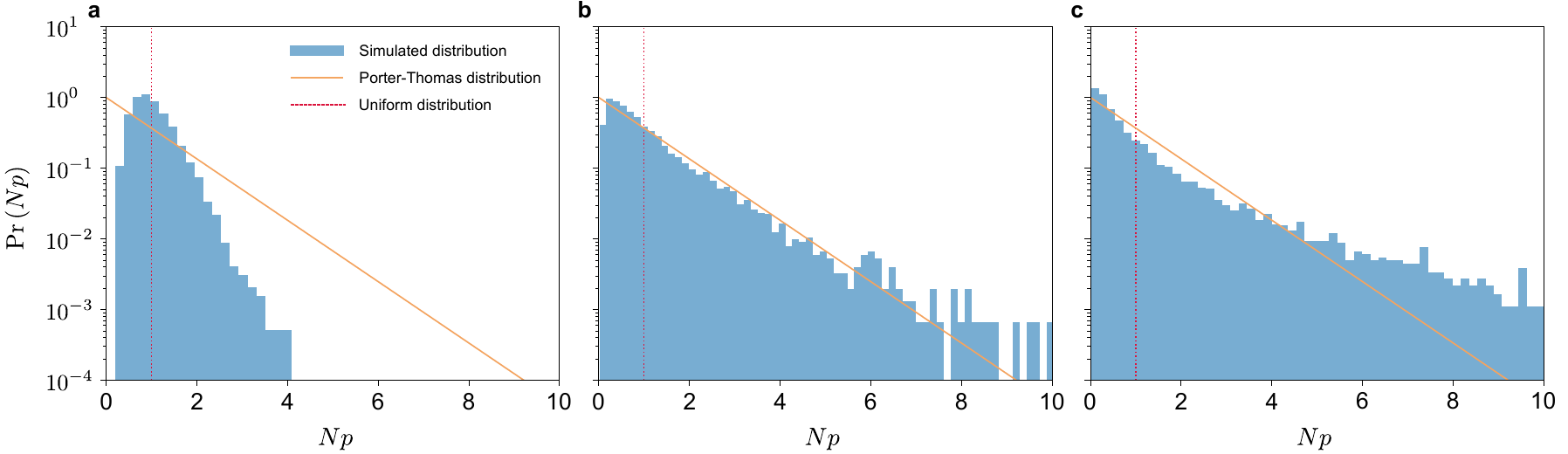}
    \caption{Output probability distribution for dynamic sampling of Haar random unitary evolution. \textbf{a - c} The plots represent the re scaled probability distributions obtained by sampling on $k$ times. Simulations comprised a 5-qubit environment interacting with a single qubit system. Sampling times vary from $k = 10$ on the left, to $k = 50$, and $k = 105$ on the right. These are compared to the orange line which represents the ideal Porter-Thomas distribution $\text{Pr}(Np) = e^{-Np}$, and the red dashed line which shows the uniform distribution $\delta(p - 1/N)$. Here, $N = 2^{k}$ is the dimension of the output bit-string of length $k$. At short sampling times the distribution is peaked around $Np = 1$, and trending towards incoherent uniform randomness. At $k = 50$ sampling times, the distribution is close to an ideal Porter-Thomas distribution, with an average minimum K-L divergence of 0.04, before pulling away at longer sampling times and demonstrating a characteristic tail suggestive of an insufficiently thermalised system.}
    \label{fig:Porter-Thomas distribution RCS}
\end{figure*}

\begin{figure}[ht]
    \centering
    \includegraphics[width=\linewidth]{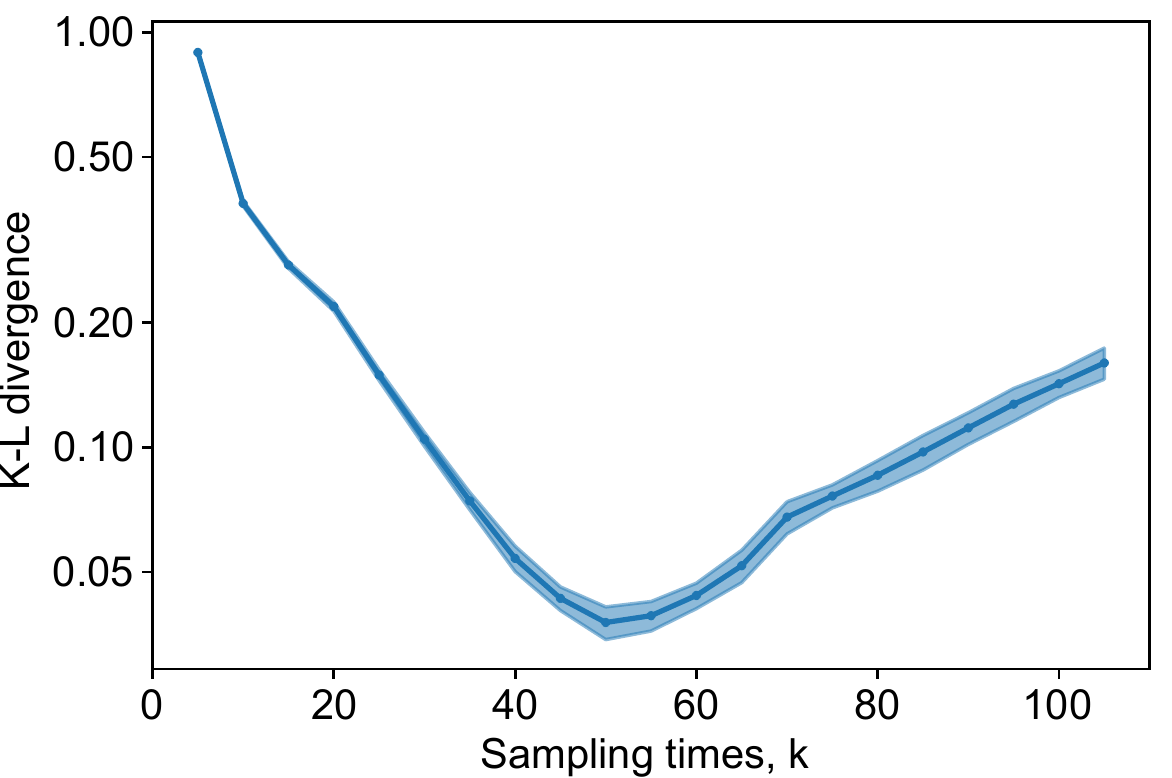}
    \caption{K-L divergence from an ideal Porter-Thomas distribution for Haar random unitary evolution. The K-L divergence for sampling times $k = 5$ to $k =105$ is shown for simulations of a 5-qubit environment interacting with a single system qubit. Each data point is the average of 30 simulations of the process for the specified sampling time. The shaded region represents the standard deviation. For each of these simulations we repeat our circuit 10,000 times to compute the probability density function. The average minimum K-L divergence is 0.0378 for sampling times $k=50$.}
    \label{fig:K-L divergence}
\end{figure}

Here, we present a series of numerical simulations of an $\mathsf{OpenDQP}$ sampling problem where the system-environment dynamics are random processes based on these models. First, we examine $SE$ evolution where the unitaries are matrices drawn from the Circular unitary ensemble (CUE) with Haar measure. Then, we consider random Hamiltonians drawn from the Gaussian unitary ensemble (GUE). 

Despite significant progress in the complexity-theoretic arguments for the hardness of RCS~\cite{bouland2022noise}, it is still unknown if sampling on a reduced output of a random circuit is hard. Not surprisingly, it is extremely challenging to prove this directly, so instead we will rely on a numerical benchmark to demonstrate classical hardness by showing that the temporal output distribution of our simulations converges to a Porter-Thomas distribution (PTD) consistent with quantum chaos~\cite{kapit2020entanglement}. 
\begin{figure*}[ht]
    \centering
    \includegraphics[width=\linewidth]{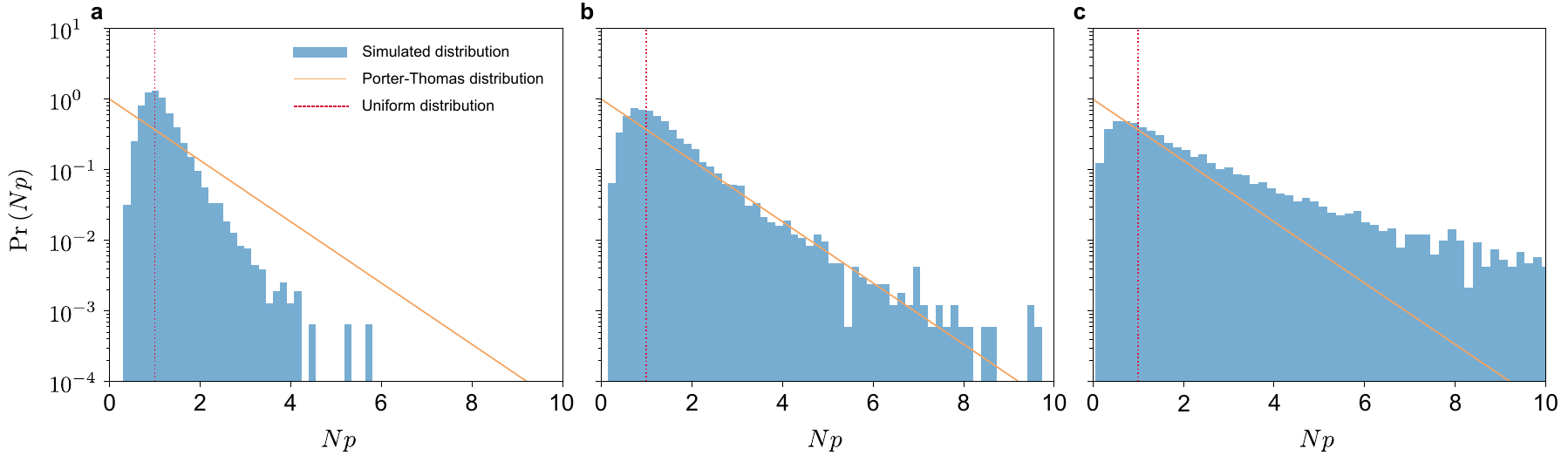}
    \caption{Output probability distribution for dynamic sampling of random Hamiltonian evolution. \textbf{a - c} The plots represent the re scaled probability distributions obtained by sampling on $k$ times. Simulations comprised a 5-qubit environment interacting with a single qubit system. Sampling times vary from $k = 5$ on the left, to $k = 15$, and $k = 35$ on the right. The orange line represents the ideal Porter-Thomas distribution $\text{Pr}(Np) = e^{-Np}$, and the red dashed line is the uniform distribution $\delta(p - 1/N)$.}
    \label{fig:Porter-Thomas distribution random Hamiltonian}
\end{figure*}

\subsection{Haar random unitary evolution}
We first describe a process with evolution based on the random circuits utilised in RCS. We consider the process tensor as shown in Figure \ref{fig:Process tensor}a, with a single system qubit interacting with an environment, of total size $n$. 
The initial system-environment state is given by $|0\rangle^{\otimes n}$, and the unitaries $U_{j:j-1}$ at each time-step are given by random matrices drawn from the Haar measure on $U(d)$, where $d = 2^{n}$ is the dimension of the matrix. 

Since the Haar measure is invariant under left and right multiplication by an independent unitary matrix, we chose our matrices randomly at each time-step so that our cumulative evolution remains Haar random. 
From a physical perspective, we can consider this process as a system and environment evolving according to a Hamiltonian that changes randomly in time. Whilst evolution according to Haar random unitaries is in itself not a physical model, and can only be achieved by random circuits in exponential time, the products of random unitary matrices appear in chaotic scattering and periodic time-dependent perturbations, which has motivated the study of the properties of composed ensembles of such unitaries~\cite{pozniak1998composed}. 

At each of the $k$ times we make a measurement in the computational basis. The results of these successive measurements form a binary bit-string of length $k$ which is sampled according to the output probability distribution given by Equation~\eqref{eq:PT_probability}. 

\subsubsection{Output statistics}
\begin{figure}[ht]
    \centering
    \includegraphics[width=\linewidth]{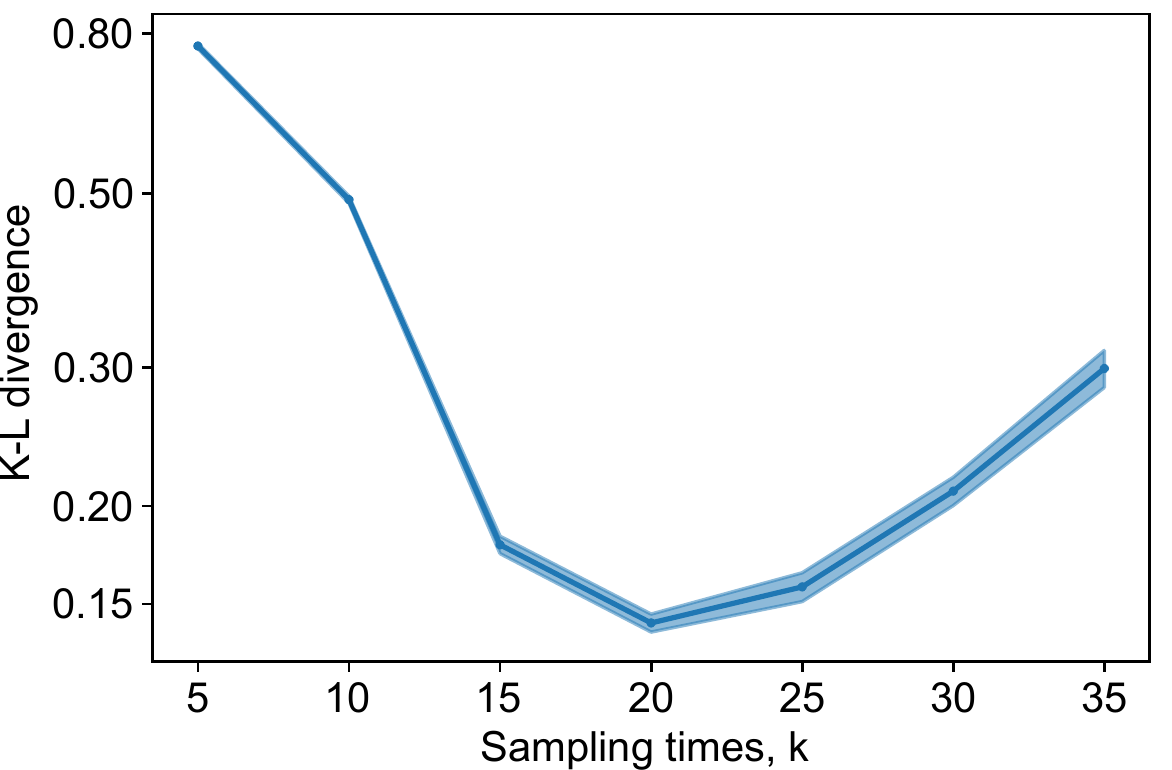}
    \caption{K-L divergence from a Porter-Thomas distribution for random Hamiltonian evolution. The K-L divergence is plotted as a function of the total number of sampling times, $k$. A minimum of 0.142 is reached at $k = 20$. The shaded region represents the standard deviation computed over 30 runs of the simulation at each sampling time.}
    \label{fig:KL divergence random Hamiltonian}
\end{figure}

We examine the output probability distribution generated by sampling on $k$ times. We consider both the shape of the distribution and the `distance' of the observed distribution to the Porter-Thomas distribution. For this, we use the Kullback-Leibler (K-L) divergence, which for two probability distributions $P$ and $Q$ is given by:
\begin{equation}
\label{eq: KL divergence}
    \mathcal{D}_{KL} (P\mathrel{\Vert}Q) \equiv \sum_{i} P_{i} \ln{ \frac{P_{i}}{Q_{i}}}
\end{equation}
The Kullback-Leibler divergence provides a measure of how well the Porter-Thomas distribution approximates our output distribution. For our calculations, $P$ is the experimentally determined distribution and $Q$ is the ideal Porter-Thomas distribution. 

In Figures~\ref{fig:Porter-Thomas distribution RCS}a - c, we plot the re-scaled output bit-string probabilities $Np$ of our simulations at varying sampling times $k$, where $N = 2^k$. For our simulations we have an environment of 5 qubits interacting with a single system qubit. We begin sampling at $k=5$ times and increase in increments of 5 until sampling on $k=105$ times. At short times (Figure~\ref{fig:Porter-Thomas distribution RCS}a), the distribution is peaked around the uniform distribution and appears to be converging towards incoherent uniform randomness. This is to be expected, since if we prepare a $n$-qubit random state and sample from only $m$ qubits then for $m \ll n$ the measured $m$-qubit state will be close to the maximally mixed state. 

\begin{figure*}[ht]
    \centering
    \includegraphics[width =\linewidth]{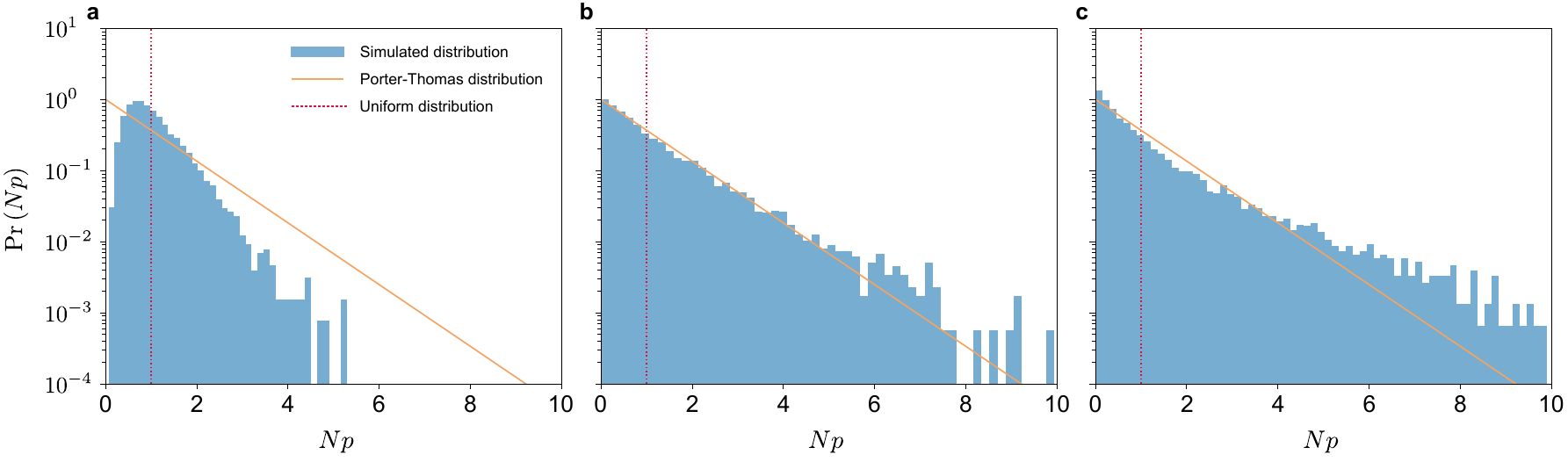}
    \caption{Output probability distribution for sampling of a random MPS state. We plot the re scaled probability distributions obtained from sampling Haar random MPS states, varying the bond dimension and number of qubits sampled. Once again, the orange line represents the ideal Porter-Thomas distribution $\text{Pr}(Np) = e^{-Np}$, and the red dashed line is the uniform distribution $\delta(p - 1/N)$. Qualitatively, these distributions follow the same transition seen in the simulation of Haar random unitary evolution and random Hamiltonian evolution for varying sampling times.
    \textbf{a} Sampling on 7 qubits of a 12-qubit Haar random MPS with maximum bond dimension of $\chi = 64$.
    \textbf{b} Sampling on all 12 qubits of a 12-qubit Haar random MPS state with maximum bond dimension.
    \textbf{c} Sampling on all 12 qubits of a 12-qubit Haar random MPS state with reduced bond dimension of $\chi = 20$.}
    \label{fig:Porter-Thomas distribution MPS}
\end{figure*}
As we increase the number of times we sample, the system gradually approaches the Porter-Thomas distribution (Figure~\ref{fig:Porter-Thomas distribution RCS}b). However, at large time steps with $k > O(2^{n})$ we observe a larger tail. These results can be confirmed by examining the K-L divergence of the output distribution from a Porter-Thomas distribution as a function of the number of sampling times, as shown in Figure~\ref{fig:K-L divergence}. The K-L divergence reaches an average minimum of approximately 0.04 when the number of sampling times is $k=50$, before the distributions diverge again at larger sampling times.

The first and most important finding is that we achieve an experimental output distribution that closely resembles a Porter-Thomas distribution when sampling on only a single qubit over time. This result provides numerical support that the RCS version of $\mathsf{OpenDQP}$ is classically hard to sample from. However, as we sample beyond the times required for convergence to a Porter-Thomas distribution our distribution stretches away and demonstrates a large tail. These tails are seen in previous work simulating random circuits~\cite{villalonga2019flexible, kapit2020entanglement} and are indicative of circuits with insufficient depth to fully thermalise. We provide further analysis of these tails in Section~\ref{sec:Tails MPS}. 

\subsection{Random Hamiltonian evolution}\label{sec: random ham}
We can utilise the same numerical tools to extend our complexity analysis to consider a more physical dynamical model, namely the evolution of a system and environment according to a random Hamiltonian. Random matrix theory has been frequently applied to the study of nuclear physics, with Hamiltonians drawn from the GUE and Gaussian orthogonal ensemble (GOE) modeling the spectral properties of nucleon excitation, particularly the breaking of time-reversal invariance in nuclei~\cite{fyodorov1997statistics}.

For our simulations we consider a simplified model of a single qubit interacting with a fixed environment of total size $n$, initiated in the state $|0\rangle^{\otimes n}$. The unitary evolution operator is now given by $U_{j:j-1} = \exp(-iH\Delta t_j)$, where $\Delta t_j$ corresponds to the time-step $t_j - t_{j-1}$, and the Hamiltonian, $H$, is a matrix selected from the GUE so that each entry is a random complex Gaussian number. The output distribution should similarly converge to a Porter-Thomas type distribution in the limit of large $N=2^n$. To make this model local, we weight the off-diagonal matrix elements $h_{ij}$ of $H$ with an exponential cut-off, i.e., $h_{ij} \mapsto \exp\{-|i-j|\} h_{ij}$. Each time-step $\Delta t_j$ is then chosen randomly as a fraction of $2\pi$. 

\subsubsection{Output statistics}
The output probability distribution for our simulations of random Hamiltonian evolution is plotted for a number of sampling times in Figures~\ref{fig:Porter-Thomas distribution random Hamiltonian}. Once again, we see a qualitative transformation in the output distribution from a distribution peaked around incoherent uniform randomness at short sampling times, to convergence to a Porter-Thomas distribution, and ultimately the emergence of tails at long sampling times. Similarly, we see a KL divergence which reaches an average minimum of 0.14 with sampling time $k = 20$, as shown in Figure~\ref{fig:KL divergence random Hamiltonian}.

Whilst ensembles of large Gaussian random matrices can be used to model the statistical properties of complex closed quantum systems, we show here that the reduced dynamics of a single qubit of these larger systems remains chaotic when sampling over time. If we assume convergence of a probability distribution to the PTD as a measure of complexity, then this is a further step towards showing that more realistic physical models of open quantum systems may be classically hard to sample from, and thus simulate.  

\subsection{Understanding the tails with random MPS}\label{sec:Tails MPS}
We now provide a heuristic analysis of the emergence of tails seen at long sampling times in our simulations. We specifically consider the Haar random evolution, although the explanation equally applies to random Hamiltonian evolution.
Since at each time-step we implement a Haar random unitary, which corresponds to a circuit of sufficient depth to demonstrate Porter-Thomas statistics on the entire system-environment state, we cannot immediately explain the tails as resulting from a non fully themalised circuit. Although we expect a similar phenomenon is occurring, and that the circuit construction of the Choi state for this process results in a circuit with insufficient depth. 

To see this, we first note that the dimension of the process tensor Choi state grows with the number of sampling times $k$. However, the bond dimension of the state will be bounded by the effective dimension of the environment. We hypothesise that as we increase our sampling times, the bond dimension of our process saturates according to the random $SE$ interaction of our unitaries, yet the dimension of our Choi state grows. Ultimately, as we sample at further times we cannot explore any more of the environment, and sampling from the resulting Choi state is akin to sampling from a random matrix product state (MPS) with non-maximal bond dimension. 

We can confirm this intuition by comparing our results in Figure~\ref{fig:Porter-Thomas distribution RCS} with the sampling of a Haar random MPS with varying bond dimension. We demonstrate this in Figures~\ref{fig:Porter-Thomas distribution MPS}a - c, where we reproduce the qualitative transition in output distribution seen in our dynamic RCS simulations. Of particular note in Figure~\ref{fig:Porter-Thomas distribution MPS}c, we are able to reproduce the tail by sampling all qubits of a 12-qubit MPS with non-maximal bond dimension $\chi = 20$.

\section{Discussion}

In this work, we initiate the study of open quantum systems from a complexity-theoretic perspective. Specifically, we examine the classical complexity of sampling an open quantum system at successive points in time. We show the existence of quantum stochastic processes that, when formulated as multi-time sampling problems of an open quantum system, are hard to classically simulate. This includes both circuit models and Hamiltonian dynamics.

The typical consideration in open quantum dynamics is to solve the corresponding master equation. 
We initially postulated that if the underlying quantum stochastic process is classically complex then a member of the associated family of master equations would also be hard -- which includes both driven and non-driven master equations. 
Our results hence provide a strong foundation for the existence of master equations that are classically hard to simulate.
In particular, given that complexity in many of our models can be controlled by system operations, then it is highly likely that many driven master equations will also be hard.
In fact our results enable us already to comment on the types of processes which will admit hard driven master equations. 
If a process is Markovian then adding a local field via unitary operations should not change the complexity of this task and the family of master equations may be efficiently simulated~\cite{jozsa2003role}. 
However, for a non-Markovian process, unitary control on the system has the potential to drastically change the complexity of simulation.

Finding complex non-driven or constant local field master equations is perhaps more challenging, since reducing the number of sampling times requires that the complexity obtained by multi-time sampling is propagated when the remaining process tensor legs are projected onto identity operations.
In itself, this task is already anticipated to be a hard problem in the worst-case, as shown by the complexity of the DQC1 model~\cite{morimae2014hardness, fujii2018impossibility} where a qubit is prepared, interacts with an infinite temperature bath, and is then sampled at some later time. 
In this setting, the set of maps is limited to those from time $t=0$ to any later time $t=t_j$, and the resulting master equation is only valid for uncorrelated initial states with time homogeneous underlying dynamics, and stationary initial environment states.
Moreover, the specific circuit used to prove the hardness of the DQC1 model is highly engineered~\cite{fujii2018impossibility}, and it is known that without entanglement the DQC1 model is easy to simulate~\cite{yoganathan2019one}.
Nonetheless, even if a single example of this restricted model is likely to be hard, then it is promising that more complicated non-driven master equations with initial correlations or inhomogeneous dynamics will also be classically hard to simulate.

To support this, 
consider the quantum stochastic process with initial system-environment state $\rho_0 = |+\rangle \langle+|_S \otimes |0\rangle   \langle 0|_E$ and Hamiltonian $H_{SE} = |0\rangle \! \langle 0|_S \otimes H_E$, where $H_E$ is drawn from the Gaussian unitary ensemble as per section~\ref{sec: random ham}. 
The complexity of the associated non-driven master equation corresponds to the set of two-time samplings within $\mathsf{OpenDQP}_2$.
Thus, we evolve our state for some time $t$ according to $U_{SE}^t = \exp\{-i H_{SE} t\} = |0\rangle\!\langle0|_S \otimes w_t + |1\rangle\!\langle1|_S \otimes \openone_{E}$ and sample the system of the resulting system-environment state $|\Psi_{SE}\rangle = \frac{1}{\sqrt{2}} \left(|0\rangle_S |w_t\rangle_E + |1\rangle_S |0\rangle_E\right)$, where $|w_t\rangle_E = w_t|0\rangle_E$. The corresponding density matrix is $\Psi_{SE} = |\Psi_{SE}\rangle \! \langle \Psi_{SE}|$, such that the reduced density matrix of the system is given by:
\begin{equation}
    \Psi_S = \frac{1}{2} \begin{bmatrix}
1 & _ E \langle 0| w_t\rangle_E \\
_ E \langle w_t| 0\rangle_E & 1
\end{bmatrix}.
\end{equation} 
Repeating this for varying values of $t$ amounts to sampling from the set of stochastic maps $\{\Lambda_{t:0}\}$.
When sampling, we can choose to measure in the $X$ and $Y$ bases with outcomes $x$ and $y$ respectively, then $\langle X\rangle^2 + \langle Y\rangle^2 = |\langle0|w_t\rangle|^2$. 
But this is precisely the probability of obtaining the all zero bit-string when sampling the environment state that has evolved according to $U_E^t = \exp\{-iH_{E}t\}$. 
And for randomly chosen $t$, the outcomes will form a probability density function with exponential shape like the Porter-Thomas distribution.
In this instance, the complexity of the master equation corresponds to sampling from the ensemble $\{|w_t\rangle\}$, which is equivalent to the ensemble of unitary matrices $\{U_E^t\}$.
In fact, for any complex $U^t$ this master equation will also be complex. 
The foundational arguments made here pave the way for determining more general master equations that are classically hard, and furthermore to delineate the subsets $\mathsf{OpenDQP}_2$ and $\mathsf{OpenDQP}_k$.

This work has bridged the gap from spatial to temporal sampling problems, which establishes the pathway to systematically examine the complexity of physically significant models of open quantum systems -- for example, understanding the paradigmatic spin-boson model. We anticipate that our results will serve as a firm foundational basis for the exploration of open quantum system simulation as an avenue to quantum advantage. Ideal application will be in understanding the complexity of important problems in the field, such as the effect of coherent driving on dissipative environments. As we have shown, quantum computers may find utility in determining features of interesting open systems: a straightforward task like finding the idle dynamics of a state, or as exotic as learning complex multi-time features in a quantum stochastic process. 

\begin{acknowledgments}
I.A.A. is supported by an Australian Government Research Training Program Scholarship, a Monash Graduate Excellence Scholarship, and the Alan P. Roberts Doctoral Scholarship.
G.A.L.W. is supported by an Australian Government Research Training Program Scholarship. 
C.D.H. is supported through a Laby Foundation grant at The University of Melbourne. 
K.M. acknowledges support from the Australian Research Council Future Fellowship FT160100073 and Discovery Project DP220101793.
K.M. and C.D.H. acknowledge the support of Australian Research Council's Discovery Project DP210100597.
K.M. and C.D.H. were recipients of the International Quantum U Tech Accelerator award by the US Air Force Research Laboratory.
\end{acknowledgments}


%

\onecolumngrid
\section*{Appendix}
\appendix
\section{The process tensor Choi state}
\label{app: Choi state details}
\begin{figure}[ht]
    \centering
    \includegraphics[width=\linewidth]{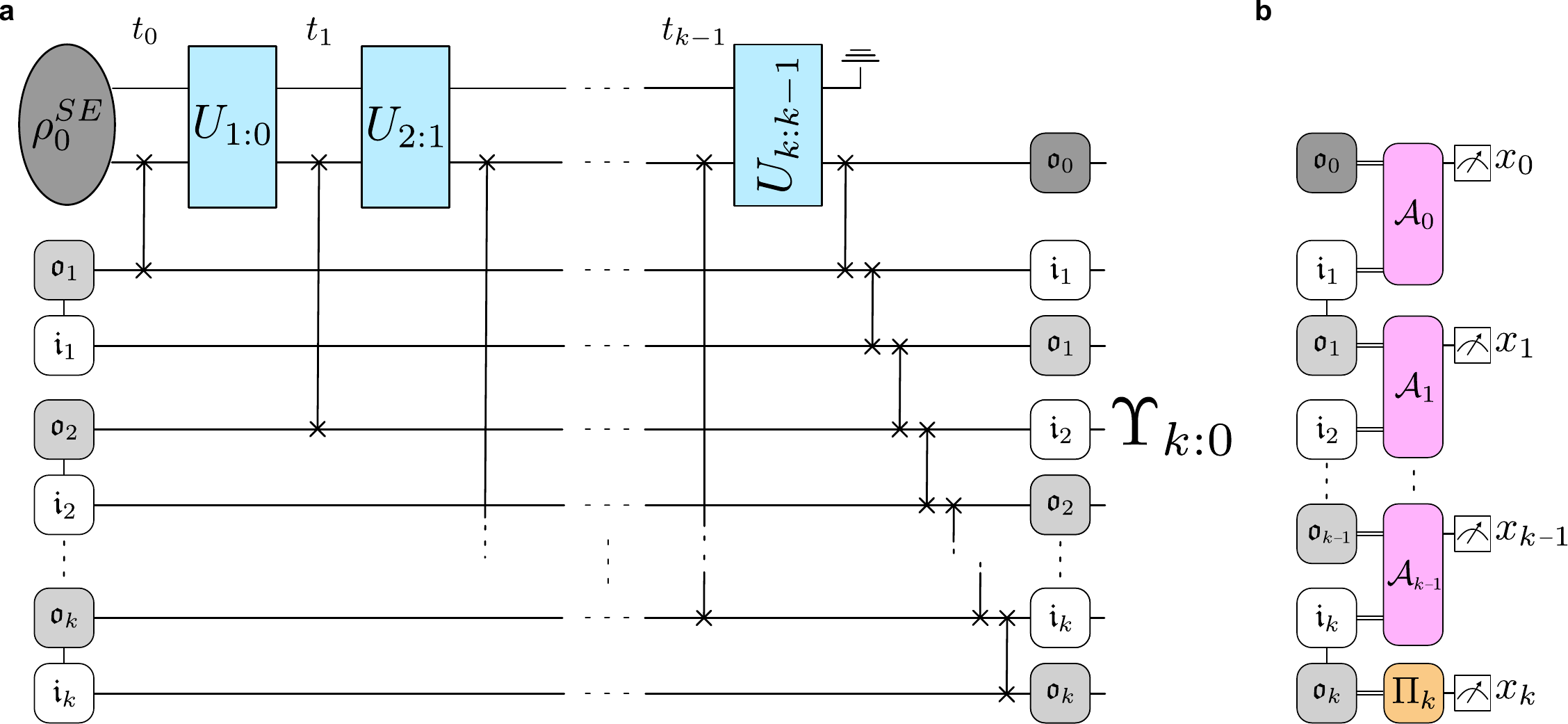}
    \caption{Sampling the process tensor Choi state.
    \textbf{a} The circuit construction of the Choi state of a $k$-step process using a generalisation of the Choi-Jamiolkowski isomorphism. One half of $k$ Bell pairs is swapped into the process at each time. At the end of the circuit a series of SWAP gates reorder the indices to correspond to causal ordering. 
    \textbf{b} The resulting Choi state can be sampled with operations $\mathcal{A}_j$ and measurement apparatus $\Pi_k$, resulting in measurement outcomes $x_j$.}
    \label{fig:Process tensor Choi state}
\end{figure}

Figure~\ref{fig:Process tensor Choi state}a provides a more detailed circuit diagram for construction of the process tensor Choi state using the generalised Choi-Jamiolkowski isomorphism. At each time, $t_j$, one half of a new Bell pair is swapped into the circuit and interacts with the environment according to the $SE$ dynamics $U_{j:j-1}$. The corresponding output leg from the process tensor, and input leg to the next time-step are labelled $\mathfrak{o}_j$ and $\mathfrak{i}_{j+1}$, respectively. The resulting state on the system and $k$ Bell pairs constitutes the $k$-step process tensor, $\Upsilon_{k:0}$, with tensor legs labeled $\{\mathfrak{o}_k,\mathfrak{i}_k,\mathfrak{o}_{k-1},\ldots,\mathfrak{i}_1,\mathfrak{o}_0\}$.

The controlled operations on the system, which the experimenter performs at each time, are represented by a set of CP maps $\{\mathcal{A}_0, \mathcal{A}_1,\ldots,\mathcal{A}_{k-1}\}$ plus a final measurement apparatus $\{\Pi_k\}$. Each operation has an associated measurement outcome $x_j$. When considering the process tensor in its Choi state representation, these operations constitute observables of the many-body state, and the resulting temporal probability distribution conditioned on the choice of instruments is found by projection onto the Choi state of the operators:

\begin{equation}
    \label{eq: Spatiotemporal Born rule detailed}
    \mathbb{P}\left(x_{k},\ldots,x_{0}|\mathcal{J}_{k},\ldots,\mathcal{J}_{0}\right) =  \text{Tr}\left[\Upsilon_{k:0}\left(\Pi_k \otimes  
    \hat{\mathcal{A}}^{(x_{k-1})}_{k-1} \otimes \cdots \otimes \hat{\mathcal{A}}^{(x_0)}_{0}\right)\right].
\end{equation}

Ultimately, the act of sampling a system sequentially over a set of times $\{t_0,t_1,\ldots,t_k\}$ has been transformed into sampling of the output of a many-body state, as shown in Figure~\ref{fig:Process tensor Choi state}b. 

\section{Sampling from the \texorpdfstring{$\mathsf{OpenDQP}\mbox{-}(\mathsf{IQP})$ process}{OpenDQP IQP}}
\label{app:pseudocode IQP}
There exists an efficient classical algorithm to sample the output of the process tensor version of $\mathsf{IQP}$. Full details of this proof are given in the text. Here, we provide the pseudocode for this algorithm.

\begin{algorithm}[H]
\caption{Sampling algorithm}
 \hspace*{\algorithmicindent} \textbf{Input}: Unitaries $\textbf{U}_{k:1}$, number of sampling times $k$, error $\epsilon$ \\
 \hspace*{\algorithmicindent} \textbf{Output}: sample $\textbf{\textup{x}}$ from distribution $\mathcal{R}(\textbf{\textup{x}})$ such that $\|\mathcal{R}(\textbf{\textup{x}}) - \mathcal{D}(\textbf{\textup{x}})_{\textbf{J}}^{k}\|_{1} \leq \epsilon$
\begin{algorithmic}[1]
\For{each $U_j$}
\State decompose $poly(n)$ 2-qubit diagonal gates into controlled-$R_z$ acting on the system, and single qubit diagonal gates on the environment
\EndFor
\State Pick an $n-1$ length bit-string uniformly at random 
\State Use the bit-string to fix the controlled-$R_z$ gates acting on the system qubit
\State Apply $poly(n)$ single qubit $R_z$ gates and $k$ measurement operators to the system qubit 
\State\Return $k$ length bit-string $\textbf{\textup{x}}$ as sample
\end{algorithmic}
\end{algorithm}

\section{Shor's algorithm}
\subsection{Choi state}
\label{app:Shor's choi state}

Here we provide a detailed construction of the process tensor Choi state for the order finding circuit of Shor's algorithm as given by the circuit in Figure~\ref{fig:Shors}b. Firstly, we consider a Choi state with indices ordered as $\{\mathfrak{o}_k,\mathfrak{o}_{k-1},\ldots,\mathfrak{o}_0,\mathfrak{i}_k,\mathfrak{i}_{k-1},\ldots,\mathfrak{i}_0\}$. This ordering is consistent with the state that results from the circuit in Figure~\ref{fig:Shors}a, where the $2n$ upper qubit register corresponds to the output legs of each Bell pair, and there are an additional $2n$ input legs corresponding to the other half of each Bell pair. We introduce the construction of the Choi state in this manner first, to familiarise the reader. Following the application of $2n$ controlled unitaries, the entire system-environment-Bell pair state is given by: 
\begin{equation}
\label{eq:Shor entire state}
    |\psi\rangle = \frac{1}{\sqrt{2^{2n}}}\sum_{x=0}^{2^{2n}-1} |xx\rangle |a^{x} \bmod N\rangle, 
\end{equation}
where n = $\log(N)$, and the modular exponential function is defined as $f(x) = a^{x}\bmod N$. Tracing over the environmental degrees of freedom gives the $k$-step process tensor Choi state:
\begin{equation}
\begin{aligned}
    \Upsilon_{k:0} &= \text{tr}_E \left( |\psi\rangle\!\langle\psi| \right)\\
    &= \frac{1}{2^{2n}} \sum_{z \in M} \langle z| \sum_{x=0}^{2^{2n}-1} \sum_{x'=0}^{2^{2n}-1} |xx\rangle\!\langle x'x'| \otimes |f(x)\rangle\!\langle f(x')| |z \rangle,
\end{aligned}
\end{equation}
where $M$ is the codomain of $f(x)$; that is, all $y$ such that $f(x) = y$. If we define $|\Psi_{i}\rangle := \sum_{x}|xx\rangle$, summed over all $x$ for which $f(x) = y_i$, then we arrive at the final representation:
\begin{equation}
\label{eq: Shor Choi state}
    \Upsilon_{k:0} = \frac{1}{2^{2n}} \sum_{i} |\Psi_{i}\rangle\!\langle \Psi_{i}|.
\end{equation}

This is the Choi state of Figure~\ref{fig:Shors}b, up to a reordering of indices. 
If we consider indices ordered according to time evolution, we can construct the associated Choi state that we project onto our controlled sequence of operations. In this setting, the final system-environment-Bell pair state is given by:
\begin{equation}
    |\psi\rangle = \frac{1}{\sqrt{2^{2n}}}\sum_{x_0 x_1 x_2 \ldots x_{2n}=0}^{1} |x_0 x_0\rangle |x_1 x_1\rangle \ldots |x_{2n} x_{2n}\rangle \otimes |a^{2^{2n} x_{2n}} a^{2^{2n-1} x_{2n-1}} \ldots a^{2^{0} x_0}\rangle.  
\end{equation}

Once again, we take the trace over the environment to retrieve the process tensor Choi state, and define $|\phi_i \rangle := \sum_{x_i} \bigotimes_i |x_i x_i \rangle$, where the summation is over all $x_i$ such that $f(x) = y_i$. The resulting Choi state can then be expressed as:
\begin{equation}
    \Upsilon_{k:0} = \frac{1}{2^{2n}} \sum_{i} |\phi_i \rangle\!\langle \phi_i | 
\end{equation}

\subsection{Sampling from the Choi state}
Sampling from the process tensor Choi state corresponds to projection of the state onto the controlled sequence of operations $\textbf{A}_{k:0}$. For the inverse QFT, the general operation at time $t_j$ is defined as:
\begin{equation}
\label{eq:Shors operation QFT}
    \mathcal{A}^{x_j}_{j} := \text{Tr}\left(M_{x_j} R_{\textbf{m}_j} \rho_j R^{\dagger}_{\textbf{m}_j}\right) |+\rangle\!\langle+|_{j+1}
\end{equation}
Here, $\textbf{m}_j:= \{m_j,m_{j-1},\ldots,m_1\}$, to indicate that the rotation operation depends on all previous measurement results. We have excluded the index $m_0$, since the first operation is simply a Hadamard gate to prepare the $|+\rangle$ state before the first controlled modular multiplication. With respect to the Choi state, we can see that the operation in Equation~\eqref{eq:Shors operation QFT} acts on legs $\{\mathfrak{o}_{j},\mathfrak{i}_{j+1}\}$. Specifically, at time $t_j$, the operation $\mathcal{A}_j$ takes in the output leg from the process $\mathfrak{o}_j$, given by state $\rho_j$, rotates it according to prior measurements, performs a measurement in the $X$-basis, before preparing the $|+\rangle_{j+1}$ state as input into the process leg $\mathfrak{i}_{j+1}$. The rotation operation $R_{\textbf{m}_j}$ is given by:
\begin{equation}
    R_{\textbf{m}_j} = \begin{bmatrix} 1 & 0 \\ 0 & \phi_j \end{bmatrix}
\end{equation}
with $\phi_j = e^{-2\pi i \sum_{k = 1}^{j-1} m_{j-k}/2^k}$.

Sampling in the computational basis then corresponds to operations $\mathcal{A}_j$, with $x_j = 0$, and thus the rotation operation simply corresponds to the identity. Using the construction of the Choi state for Shor's algorithm we can provide an alternative proof demonstrating that it is classically easy to sample in the computational basis at each time of the process. Consider the state in Equation~\eqref{eq: Shor Choi state}.  Sampling in the computational basis on the resulting process $\Upsilon_{k:0}$ then corresponds to measuring the output legs $\{\mathfrak{o}_k,\mathfrak{o}_{k-1},\ldots,\mathfrak{o}_0\}$ of this state, and projecting the input legs $\{\mathfrak{i}_k,\mathfrak{i}_{k-1},\ldots,\mathfrak{i}_0\}$ onto the post measurement state, which without loss of generality we can choose to be $|0\rangle \! \langle0|$. This is, therefore, equivalent to tracing the lower register of $n$-qubits in Figure~\ref{fig:Shors}a, and measuring the upper register of $2n$-qubits before the inverse QFT is performed. For simplicity, we consider the complexity of this equivalent sampling problem.   
To simulate the outcome of this sampling problem, we proceed as follows. The function $f$ is periodic with period $r$, and the number of periods is therefore given by $A = 2^{2n}/r$. Taking the partial trace over the environment amounts to measuring the environment and discarding the result. Since each possible value of the function $f$ occurs the same number of times, precisely $A$ times, we can select one outcome, $y_0$, uniformly at random. This projects the register of $2n$-qubits into an equal superposition of the total $A$ values of $x$ such that $f(x) = y_0$. The resulting state is therefore:
\begin{equation}
    |\psi_{2n}\rangle = \frac{1}{\sqrt{A}} \sum_{j=0}^{A-1} |x_0 + jr\rangle 
\end{equation}
Measuring this register then gives a bit-string $x_0 + j_0 r$, where once again each value of $j_0$ is equally likely and so is chosen uniformly at random. Overall, this returns a number between $0$ and $2^{2n}$ uniformly at random.

\end{document}